\documentclass[aps,prd,nofootinbib,12pt,reprint,longbibliography,superscriptaddress,floatfix,floats,tightenlines]{revtex4-2}

\usepackage{graphicx}
\usepackage{epstopdf}
\usepackage{amsmath}
\allowdisplaybreaks
\usepackage{amssymb}
\usepackage{amsfonts}
\usepackage{amsthm}
\usepackage{color}
\usepackage{array}
\usepackage{verbatim}
\usepackage{url}
\usepackage{graphicx}
\usepackage{braket}
\usepackage{epsfig}
\usepackage[
      colorlinks=true,
      linkcolor=blue,
      urlcolor=magenta,
      filecolor=blue,
      citecolor=red,
      pdfstartview=FitV,
      pdftitle={},
      pdfauthor={},
      pdfsubject={},
      pdfkeywords={},
      pdfpagemode={},
      bookmarksopen=true
	  ]{hyperref}
\usepackage{orcidlink}
\usepackage{subfigure}
\usepackage[normalem]{ulem} 
\usepackage{cancel} 
 
\usepackage{soul} 

\def\frac#1#2{{#1\over #2}}

\def\be{\begin{equation}}
\def\ee{\end{equation}}
\def\ba{\begin{eqnarray}}
\def\ea{\end{eqnarray}}

\definecolor{armygreen}{rgb}{0.29, 0.33, 0.13}
\definecolor{darkspringgreen}{rgb}{0.09, 0.45, 0.27}

\numberwithin{equation}{section}

\begin{document}

\title{Blackhole perturbations in the Modified Generalized Chaplygin Gas model }

 \author{Sunil Singh Bohra\orcidlink{0009-0007-8257-9245}}
 \email{sunilsinghbohra87@gmail.com}
 \affiliation{Center for Theoretical Physics, Jamia Millia Islamia, New Delhi - 110025, India}


\date{\today}
\begin{abstract}
We study blackhole in the Modified Generalized Chaplygin Gas (MGCG) model, a modified gravity framework proposed to unify dark matter and dark energy. The resulting spacetime is asymptotically non-flat and feature two distinct horizons, determined by the Chaplygin parameter $\alpha$. Observational constraints favor negative values of $\alpha$, which restrict the allowed parameter space. Through linear perturbation analysis, we show that this blackhole remain stable under both scalar and electromagnetic perturbation. The corresponding quasinormal mode spectra depend sensitively on the MGCG parameters, indicating that this cosmological model leaves characteristic imprints on blackhole oscillations. These results suggest that gravitational-wave observations of blackhole ringdowns may provide a means of constraining MGCG and testing unified dark sector scenarios in the strong-field regime.
\end{abstract}

\maketitle
\section{Introduction}

The rotational velocity of galaxies is expected to decline with distance from the center, reflecting the distribution of visible matter. However, observations show flat rotation curves at large radii, indicating an unseen mass component~\cite{Rubin1980}. Gravitational lensing also reveals more total mass than can be explained by baryonic matter alone \cite{Clowe2006}. These findings, together with mass discrepancies in galaxy clusters \cite{Zwicky1933} and high velocity dispersions in dwarf galaxies \cite{Walker2009}, point to the presence of dark matter.

In contrast, Type Ia supernovae, treated as standardizable candles, provide measurements of the luminosity distance $D_L(z)$ as a function of redshift\cite{Coelho:2014lba,Holz:2004xx}. Distant supernovae appear dimmer than expected in a decelerating universe, indicating late-time acceleration \cite{Riess1998,Perlmutter1999}. Since $D_L(z)$ depends only on the expansion history, this offers strong geometric evidence for dark energy. Additional support comes from strong lensing systems, where time delays between multiple quasar images yield the time-delay distance $D_{\Delta t}$, which constrains the expansion rate, especially $H_0$, when the lens mass distribution is well modeled~\cite{Suyu2017}. Further confirmation arises from cosmic chronometers, which estimate $H(z)$ by comparing the differential ages of passively evolving galaxies at different redshifts. These measurements, when interpreted within the $\Lambda$CDM framework, suggest a transition from deceleration to acceleration~\cite{Moresco2016}.

While dark matter and dark energy are usually treated as distinct components, their gravitational effects have inspired unified models that describe both within a single framework. The Generalized Chaplygin Gas (GCG) model \cite{Kamenshchik2001,Bento2002} is one such approach, introducing an exotic fluid that behaves like pressureless matter at early times and like dark energy at late times. This makes the GCG a compelling candidate for modeling the entire dark sector in a unified way.

Although the GCG model offers a unified description of the dark sector, it faces difficulties when matched with observations. Notably, it struggles to reproduce the cosmic microwave background (CMB) and large-scale structure data due to oscillations in the matter power spectrum \cite{Sandvik2004}. These issues are linked to its non-zero sound speed, which affects the growth of density perturbations \cite{Amendola2003,Bean2003}. To overcome these limitations, extensions like the Modified Chaplygin Gas (MCG) \cite{AnandSen2004} and Modified Generalized Chaplygin Gas (MGCG) \cite{Lu2008,Xu2011,Feng2008} have been developed for better agreement with data.

In the Modified Generalized Chaplygin Gas (MGCG) model, late-time acceleration emerges from modifications to the background geometry rather than exotic matter content, making MGCG a promising candidate for unified dark sector cosmology \cite{AnandSen2004,Lu2008,BouhmadiLopez2005}. Such modifications affect not only the cosmic background evolution but also the perturbative dynamics of compact objects. This motivates the study of quasinormal modes (QNMs) in MGCG-based spacetimes.

Quasi-Normal Modes (QNMs) are a fundamental feature of blackhole perturbation theory, representing the characteristic oscillations that dominate the late-time behavior of a perturbed compact object. The theoretical foundation for QNMs was established in the mid-20th century through early studies of blackhole stability.

The first rigorous treatment of perturbations around a Schwarzschild blackhole was provided by Regge and Wheeler in 1957~\cite{PhysRev.108.1063}, who studied axial perturbations and demonstrated the stability of the Schwarzschild metric. This was extended by Zerilli in 1969~\cite{PhysRevD.2.2141}, who developed a formalism for polar perturbations. These works showed that both axial and polar perturbations satisfy Schrödinger-like wave equations with effective potentials, setting the stage for mode analysis.

The physical interpretation of QNMs was introduced by Vishveshwara~\cite{VISHVESHWARA:1970}, who demonstrated that a perturbed blackhole exhibits damped oscillations-what we now identify as QNMs. His work revealed that these modes encode the response of a blackhole to external perturbations. The study of rotating blackholes followed soon after, with Press and Teukolsky~\cite{Press:1973zz,Teukolsky:1973ha,Teukolsky:1974yv} deriving the perturbation equations for the Kerr geometry. Their work also led to the description of superradiant scattering, later expanded by Starobinsky~\cite{1973Starobinsky}.

Subsequent decades saw the development of computational and analytical methods for determining QNM frequencies. Chandrasekhar and Detweiler~\cite{Chandrasekhar:1975nkd} were the first to numerically compute QNM spectra, while Leaver~\cite{Leaver:1985} introduced the continued fraction method to solve the master equation governing blackhole perturbations with high precision.

Analytical approaches were also explored. Schutz and Will~\cite{Schutz:1985km} applied the WKB approximation to the blackhole potential, interpreting it as a quantum mechanical barrier penetration problem. This method was later extended to higher orders by Konoplya\cite{Konoplya:2003ii,Konoplya:2020hlu}, improving accuracy and broadening its applicability.

In cosmological contexts, Moss and Norman~\cite{Mellor:1989ac,norman.Moss:2001ga} analyzed perturbations in de Sitter spacetimes, establishing the role of boundary conditions in QNM behavior. Brady et al.~\cite{Brady:1999wd,brady:krivan:1997pclp} conducted time-domain studies of Schwarzschild-de Sitter blackholes, observing late-time exponential decay distinct from asymptotically flat scenarios. Cardoso and Lemos~\cite{Cardoso_2003} investigated perturbations near extremality, where the effective potential approaches the solvable Pöschl–Teller form. Konoplya’s WKB extension to higher orders allowed precise computation of QNMs for various spin fields, and Zhidenko~\cite{Zhidenko:2003wq} showed that in near-extremal SdS spacetimes, the QNM spectrum depends primarily on the blackhole mass and the cosmological constant.

These foundational works have established QNMs as a cornerstone in blackhole physics and gravitational wave astrophysics, providing both theoretical insight and observational signatures that continue to guide modern research.
Recent advances in gravitational-wave detection have renewed interest in QNMs as probes of fundamental gravity. In general relativity, axial and polar perturbations of Schwarzschild blackholes are isospectral, but this property is typically broken in modified gravity, where additional degrees of freedom can generate new families of modes sensitive to the underlying theory \cite{berti2015testing,tattersall2018quasinormal}. Distinct QNM spectra have also been predicted for exotic compact objects and alternative spacetime structures, including gravastars, wormholes, and blackholes in extended gravity, offering potential observational signatures beyond general relativity \cite{Konoplya:2016hmd,Pani:2012bp}. Quasinormal modes and greybody factors of black holes surrounded by modified Chaplygin gas have likewise been studied, showing that the gas parameters can significantly influence stability and radiation \cite{Sekhmani2025}.

The detection of multiple QNMs in gravitational-wave signals enables blackhole spectroscopy, where resonant frequencies are used to infer mass and spin and to test the no-hair conjecture by checking consistency with Kerr predictions \cite{isi2019testing,Berti:2005ys}. These developments highlight the role of QNMs as both theoretical tools and observational probes in the strong-field regime of gravity. The paper is organized as follows. Section~\ref{Blackhole} describes the structure of the horizon and explores the parameter space. Section~\ref{QNM} highlights the importance of QNMs as field perturbations. Section~\ref{perturbation} explains the procedure of treating test fields as scalar and electromagnetic perturbations. Section~\ref{effective pot} correlates the effective potential with the QNMs and the horizon structure, along with the applicability of the WKB method. Section~\ref{WKB} outlines the method used to solve the differential equation. Finally, Section~\ref{result} discusses the effects of modified gravity on the QNMs.

\section{Blackhole in Modified generalised  Chaplyin  gas model}\label{Blackhole}
The generalized Chaplygin gas (GCG) model \cite{Kamenshchik:2001cp,Bento:2002ps,Bento:2004uh} provides a unified description of dark matter and dark energy by introducing an exotic fluid into the matter sector of FRW cosmology. An alternative approach, which uses only dust content yet reproduces the key features of GCG cosmology, involves modifying gravity itself \cite{Barreiro:2004bd}. The spherically symmetric spacetime in this modified Generalized  Chaplygin gas (MGCG) framework is described by  

\begin{equation}
 ds_{g}^{2}  = g_{00}dt^{2}-g_{11}dr^{2}-r^{2}\left(d\theta^{2}+\sin^{2} \theta d \phi^{2}\right)
\end{equation}
 where
\begin{equation}
g_{00} =  g_{11}^{-1} = 1-r^{2} H_{0}^{2}  g \left(\frac{r_{c}^{3}}{r^{3}}\right); 
\end{equation}
and
\begin{equation} 
g(x) = \left(1-\Omega_{m}^{\alpha +1} + x^{\alpha +1}\right)^{\frac{1}{\alpha+1}} 
\end{equation}

Here, $x$ is a dimensionless quantity. $\Omega_{m}$ is the matter energy density in the GCG model, which is scaled to lie between 0 and 1, but in the MGCG  it is treated as a parameter of the model. The parameter $\alpha$ is strictly positive in the GCG model, but in the MGCG framework it can take both positive and negative values. The MGCG spacetime model therefore contains two parameters and two constants.  

By a coordinate transformation and using the invariance of the metric length, the term $r_{c}^{3} H_{0}^{2}$ can be scaled to unity.  

In the limit $x\gg1$, subleading terms in $g(x)$ become negligible, and the metric reduces exactly to the Schwarzschild form. In this regime, the relationship between $r_{c}$ and $M$ can be established by comparison with the Schwarzschild metric. The parameter $r_{c}$ determines the length scale over which modifications to general relativity become significant \cite{Multamaki:2006zb}. For certain ranges of parameter space, the horizon condition is satisfied. The condition $g_{11}^{-1} = 0$ signals the presence of a horizon \cite{Faraoni:2020,Visser:2004,Giuliani:2024} if it holds within a specific region of parameter space.

\begin{equation}
1-r^{2} H_{0}^{2}\left(1-\Omega_{m}^{\alpha +1} +  \left(\frac{r_{c}^{3}}{r^{3}}\right)^{\alpha +1}\right)^{\frac{1}{\alpha+1}} =0 
\end{equation} 
is simplified as 
$$ 1-\Omega_{m}^{\alpha +1} + \left(\frac{r_{c}^{3}}{r^{3}}\right)^{\alpha +1} = \left(\frac{1}{r^{2} H_{0}^{2}}\right)^{\alpha +1} $$
 which finally reduces to 
\begin{equation}
H_{0}^{2(\alpha+1)}(1-\Omega_{m}^{\alpha +1}) r^{3(\alpha+1)}- r^{(\alpha+1)} +( r_{c}^{3} H_{0}^{2})^{\alpha+1}=0 
\end{equation}  

The resulting expression is an algebraic equation in $r$. Applying the coordinate transformation 
$r^{\alpha+1} = u$ reduces the equation to a cubic in $u$:  

\begin{equation}
H_{0}^{2(\alpha+1)}(1-\Omega_{m}^{\alpha +1}) u^{3}- u +( r_{c}^{3} H_{0}^{2})^{\alpha+1}=0 
\end{equation}
or 
\begin{equation} \label{horizon equation}
u^{3}-\left(\frac{1}{H_{0}^{2(\alpha+1)}(1-\Omega_{m}^{\alpha +1})}\right) u + \frac{r_{c}^{3(\alpha+1)}}{1-\Omega_{m}^{\alpha +1}}=0
\end{equation}

To determine the nature of the roots, the discriminant ($\Delta$) of the equation is analyzed:  

\begin{equation}
\Delta =\frac{4}{(1-\Omega_{m}^{\alpha +1})^{3}}\left(\frac{1}{H_{0}^{2}}\right)^{3(\alpha+1)}-\frac{27}{(1-\Omega_{m}^{\alpha +1})^{2}} \left({r_{c}^{3}}\right)^{2(\alpha+1)} 
\end{equation}

The equation admits three real roots for $\Delta > 0$, with at least one of them necessarily negative. This restricts the parameter space such that the parameters must satisfy the condition:  

\begin{equation}
1-\Omega_{m}^{\alpha +1} < \frac{4}{27}\left(\frac{1}{r_{c} H_{0}}\right)^{6(\alpha+1)}
\end{equation}  

Solving eq.~(\ref{horizon equation}) in this restricted region yields three real roots - two positive and one negative. However, the algebraic solution involves complex terms, despite all roots and parameters being real. This situation, known as \textit{casus irreducibilis} \cite{stewart2015galois}, arises when a cubic equation with three real roots is solved using radicals \cite{gullberg1997mathematics}. Although the roots are real, the solution unavoidably passes through complex intermediate steps. To express the roots in a purely real form, the trigonometric identity  

\begin{equation}
\cos{3\theta}= 4\cos^{3}{\theta}- 3\cos{\theta}
\end{equation}

is used. Now, transform the variable $u$ in eq.~(\ref{horizon equation}) as $u = v \cos\theta$. Choose the constant $v$ such that the coefficients of $\cos 3\theta$ and $\cos\theta$ in the resulting expression match those in the standard trigonometric identity. By comparing the two forms \cite{Nickalls:1993}, the following relation is obtained:  

\begin{equation}
\cos{3\theta} = \frac{3}{2} \left(r_{c}^{3}H_{0}^{2}\right)^{\alpha +1} \sqrt{3 H_{0}^{2(\alpha+1)} \left(1-\Omega_{m}^{\alpha +1}\right)}
\end{equation}

Substituting the value of $\theta$ into the transformed variable yields three distinct roots, corresponding to the three values of $n$. Among these, one is a negative real root, while the remaining two are positive real roots.  

\begin{equation}
u = \frac{2}{\sqrt{3}}\frac{1}{H_{0}^{(\alpha+1)}} \sqrt{ \frac{1}{\left(1-\Omega_{m}^{\alpha +1}\right)}} \cos{\theta}
\end{equation}
where 
\begin{equation}
\theta =\frac{1}{3} \cos^{-1}{\left( \frac{3}{2} \left(r_{c}^{3}H_{0}^{3}\right)^{\alpha +1} \sqrt{3 \left(1-\Omega_{m}^{\alpha +1}\right)}\right)}- \frac{2\pi n}{3} 
\end{equation}

for $n=0,1,2$.  

The case $n = 0$ corresponds to the larger positive root, which is identified as the cosmological horizon:  

\begin{equation}
u_{c} = \frac{2}{\sqrt{3}}\frac{1}{H_{0}^{(\alpha+1)}} \sqrt{ \frac{1}{\left(1-\Omega_{m}^{\alpha +1}\right)}} \cos{\frac{\phi}{3}}
\end{equation}

While $n = 1$ corresponds to the smaller positive root, which represents the event horizon of the black hole:  

\begin{equation}
u_{h} = \frac{2}{\sqrt{3}}\frac{1}{H_{0}^{(\alpha+1)}} \sqrt{ \frac{1}{\left(1-\Omega_{m}^{\alpha +1}\right)}} \cos\left({\frac{\phi}{3} } - \frac{2\pi}{3}\right)
\end{equation}

where  

\begin{equation}
\phi= \cos^{-1}{\left( \frac{3}{2} \left(r_{c}^{3}H_{0}^{3}\right)^{\alpha +1} \sqrt{3 \left(1-\Omega_{m}^{\alpha +1}\right)}\right)}
\end{equation}

The region between the event and cosmological horizons is causally connected. Beyond the cosmological horizon, no physical interaction is possible.  

\begin{figure}[tpb]
	\centering
	\includegraphics[width=0.85\linewidth]{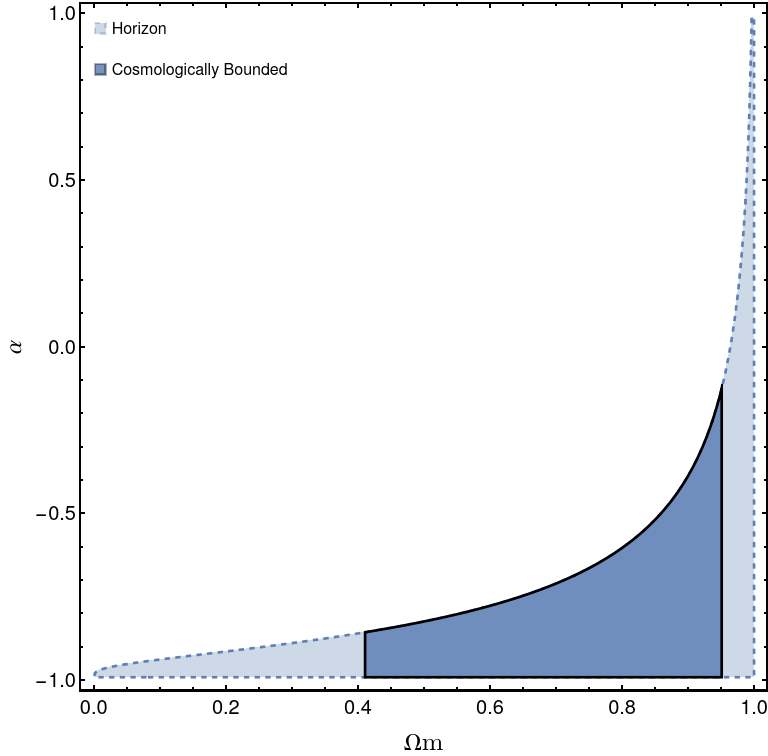}
	\caption{Parameter space showing allowed region under the MGCG blackhole horizon condition (light-shaded). The dark blue region, bounded by a closed thick black line (three straight edges and one curved), satisfies additional cosmological constraints. The curved boundary region corresponds to the near-extremal regime.}
	\label{fig:parameter_space}
\end{figure}

In the MGCG-based blackhole model, there are two parameters, Fig.~(\ref{fig:parameter_space}). Theoretically, $\Omega_{m}$ ranges from 0 to 1, while $\alpha$ can take any value, either positive or negative  \cite{Sen:2005sk}. However, requiring the model to describe a blackhole with a well-defined horizon structure imposes constraints on the parameter space. Additional restrictions arise from observational data related to the parameters of the Modified Chaplygin Gas model. We have explored the $2\sigma$ or $95 \% $ confidence region around the experimental values of these parameters. Both sets of constraints point toward negative values of the $\alpha$ parameter.  

One of the key advantages of the Modified Chaplygin Gas Cosmology (MCGC) framework emerges in comparison with the Schwarzschild-de Sitter case, which corresponds to $\alpha = 0$. In that limiting case, the parameter $\lambda$ is restricted to values below $0.11$, which implies a high lower limit on the matter density parameter: $\Omega_{m} \geq 0.963$. However, when $\alpha$ takes non-zero values-as allowed in the MGCG scenario-this condition is significantly eased. The additional flexibility introduced by non-zero $\alpha$ permits a wider and more realistic range of $\Omega_{m}$, making MGCG a more adaptable and observationally consistent model for describing late-time cosmic acceleration.

\section{Quasi-Normal Modes}\label{QNM}

Quasi-Normal Modes (QNMs) are the eigenmodes of dissipative systems and arises, in the perturbed black holes. The dynamics of the blackholes are governed by a few physical parameters-such as mass, spin and additional parameter of the modified gravity theories.~\cite{Abbott:2016apu}.

Astrophysical events capable of disturbing spacetime typically involve strong gravitational fields. Blackholes, neutron stars, and other compact objects are expected sources. Among them, blackholes are particularly hard to observe directly due to their lack of electromagnetic emission and great distances. Consequently, QNMs are especially valuable, as they encode black hole properties in gravitational wave signals.

The ringdown signal after a merger is well described as a superposition of QNMs of the remnant blackhole. In the linear regime, these correspond to discrete complex frequencies set by the black hole’s geometry~\cite{berti2009quasinormal}. Fourier analysis reveals no frequencies beyond the QNM spectrum~\cite{dreyer2004black}. In the nonlinear regime, arbitrary frequencies still do not appear; instead, structured features like overtones and mode couplings arise, constrained by the system’s fundamental properties~\cite{giesler2019black,london2014modeling}.

The discrete nature of the QNM spectrum arises from boundary conditions requiring ingoing waves at the horizon and outgoing waves at infinity~\cite{yang2015quasinormal}. The event horizon thus plays a key role in determining the spectrum~\cite{baibhav2020multimode,isi2019testing}.

The evolution of a perturbation typically proceeds through three stages: an initial burst, a QNM-dominated ringing phase, and a late-time decay. In asymptotically flat spacetimes, this decay follows a power-law tail, while in asymptotically non-flat geometries, it is exponential~\cite{Suneeta:2003bj}. The QNM phase is largely independent of the initial conditions and is governed solely by the black hole parameters.

In the eikonal limit, where the multipole number satisfies $\ell \gg 1$, the QNM frequencies derived from the WKB method admit a geometric interpretation, in which the real part corresponds to the angular frequency of unstable circular null geodesics, while the imaginary part is related to the Lyapunov exponent characterizing their instability~\cite{Cardoso2009}.

A standard scenario for QNM excitation is the merger of two blackholes. The remnant is initially perturbed and emits gravitational radiation dominated by QNMs. The damping time is determined by the imaginary part of the QNM frequency, which depends on the remnant’s mass, spin, and any additional parameters relevant to the gravitational theory. These features allow precise extraction of physical parameters from post-merger gravitational wave data~\cite{Abbott:2016apu}.

The presence of a cosmological constant modifies the QNM structure through changes in boundary conditions. In Schwarzschild-de Sitter (SdS) spacetimes, the cosmological horizon leads to exponential decay at late times, in contrast to the power-law tails in asymptotically flat cases~\cite{Brady:1999wd,Mellor:1989ac}. For the $\ell = 0$ mode, the decay rate asymptotically approaches a nonzero constant. In Anti-de Sitter (AdS) spacetimes, reflective boundaries can produce purely imaginary frequencies for certain field configurations~\cite{cadoso.lemo.2001ads}. In near-extremal SdS blackholes-where the event and cosmological horizons nearly coincide-the QNM spectrum becomes independent of the spin of the perturbing field and depends primarily on the blackhole mass and the cosmological constant~\cite{Zhidenko:2003wq}.

 \section{ perturbation} \label{perturbation}
     \subsection{Scalar field}
      A test scalar field is considered in the background of a blackhole. The dynamics of this field are significantly influenced by the curvature of spacetime in the vicinity of the blackhole. The blackhole's characteristics, encoded in the spacetime metric, play an important and distinctive role in governing the field's evolution. By analyzing the scalar field dynamics, one can infer properties of the underlying blackhole geometry. In this work, we restrict the analysis to first-order perturbations. Since the field is treated as a test field, it does not contribute to the spacetime curvature\cite{barack2009}. This assumption is justified because the impact of second-order terms on physical observables is generally negligible compared to first-order contributions, which are sufficient to qualitatively assess the stability of the black hole.
 
     The equation governing the dynamics of a scalar field $\phi$ in a curved spacetime is the covariant Klein-Gordon equation \cite{birrell_davies, wald_gr}:
\begin{equation}
   \frac{1}{\sqrt{-g}} \, \partial_\mu \left( \sqrt{-g} \, g^{\mu\nu} \partial_\nu \phi \right) - m^2 \phi = 0,
\end{equation}
where $m$ is the mass of the scalar field, and $g$ denotes the determinant of the metric tensor.

For the spherically symmetric spacetime arising in the Modified Chaplygin Gas (MCG) model, the determinant of the metric is given by $g = -r^4 \sin^2\theta$.

Exploiting the spacetime symmetries, such as time translation invariance and spherical symmetry, the scalar field can be decomposed as \cite{mukhanov_physics, berti_qnm_review}:

\begin{equation}
    \phi(t, r, \theta, \varphi) = e^{-i \omega t} \, Y_{lm}(\theta, \varphi) \, \frac{\psi(r)}{r},
\end{equation}
where $Y_{lm}(\theta, \varphi)$ are the spherical harmonics, and $\omega$ is the frequency of the mode.

Substituting this ansatz into the Klein-Gordon equation and simplifying yields a radial differential equation of the Schrödinger-like form. Performing the radial coordinate transformation $r^{\alpha + 1} = u$ transforms the master equation into:
\begin{equation}
    \frac{d^2 \Psi(u_{*})}{du_*^2} + \left[ \omega^2 - V_{\text{eff}}(u) \right] \Psi(u_*) = 0,
\end{equation} 
with $u_{*}$ is defined as 
$$ \frac{du_{*}}{du} = \frac{1}{f(u)}$$
where the effective potential $V_{\text{eff}}(u)$ is given by:
\begin{equation}
    V_{\text{eff}}(u) = f(u) \left[ \frac{\ell(\ell+1)}{u^{\frac{2}{\alpha+1}}} + \frac{u}{u^{\frac{2}{\alpha+1}}} \frac{df}{du} + m^2 \right],
\end{equation}
and the function $f(u)$ is defined as:
\begin{equation}
    f(u) = 1 - c_2 u^{-\frac{1}{1 + \alpha}} \left(c_1 u^3 + 1\right)^{\frac{1}{1 + \alpha}},
\end{equation}
with constants:
\begin{equation}
    c_1 = \frac{1 - \Omega_m^{\alpha+1}}{r_c^{3(\alpha+1)}}, \qquad c_2 = r_c^3 H_0^2.
\end{equation}
This potential acts as a scattering barrier for the scalar field, capturing the spacetime's influence on wave propagation. The power of the new coordinate lies in its ability to compress a vast radial distance-such as the range from $r \simeq 2.11$ to $r \simeq 10^{25}$-into a much smaller domain, specifically $u \simeq 1.03$ to $u  \simeq 19.22$. This compactification significantly improves numerical resolution in both the near-horizon region and the asymptotic regime, enabling more precise analysis of scalar field dynamics across the entire spacetime. Importantly, while the coordinate transformation alters the location of the potential peak, its value remains unchanged. The peak simply shifts to a new position consistent with the transformed coordinate system.

 \begin{figure*}[htb!]
	\centering
	\minipage{0.33\textwidth}
	\includegraphics[width=\linewidth]{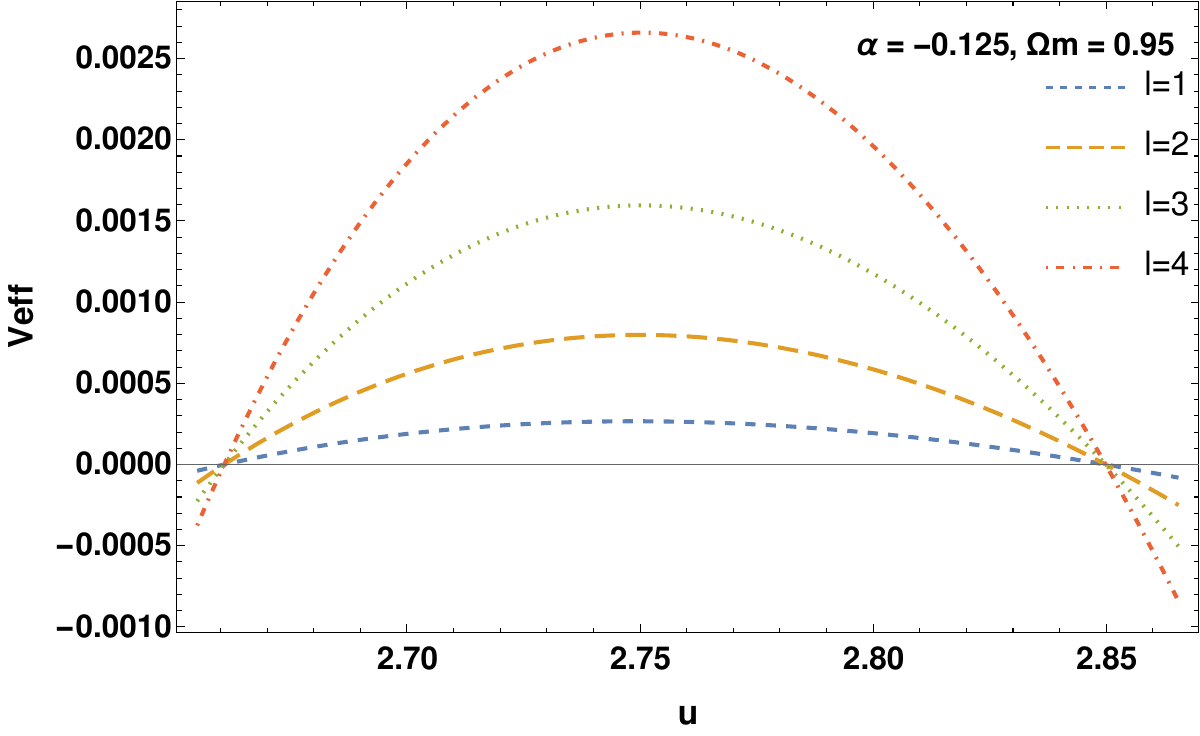}
	\endminipage\hfill
	\minipage{0.33\textwidth}
	\includegraphics[width=\linewidth]{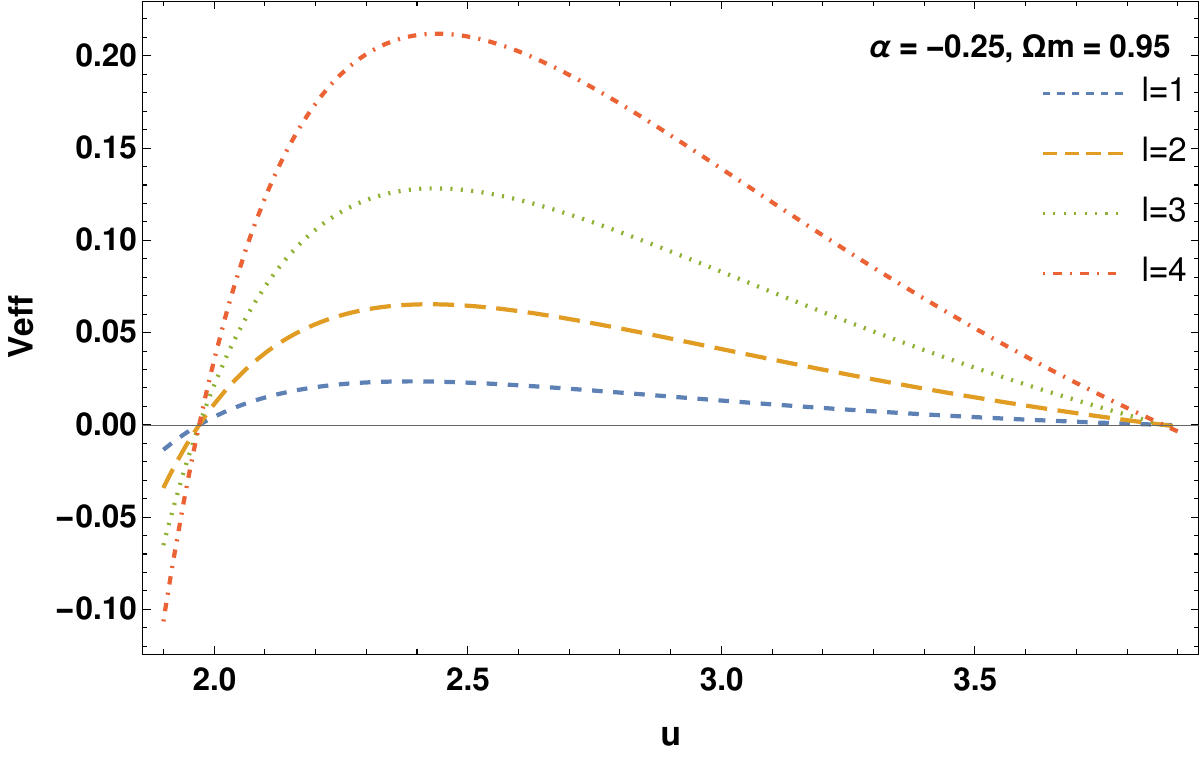}
	\endminipage
 \hfill
	\minipage{0.33\textwidth}
	\includegraphics[width=\linewidth]{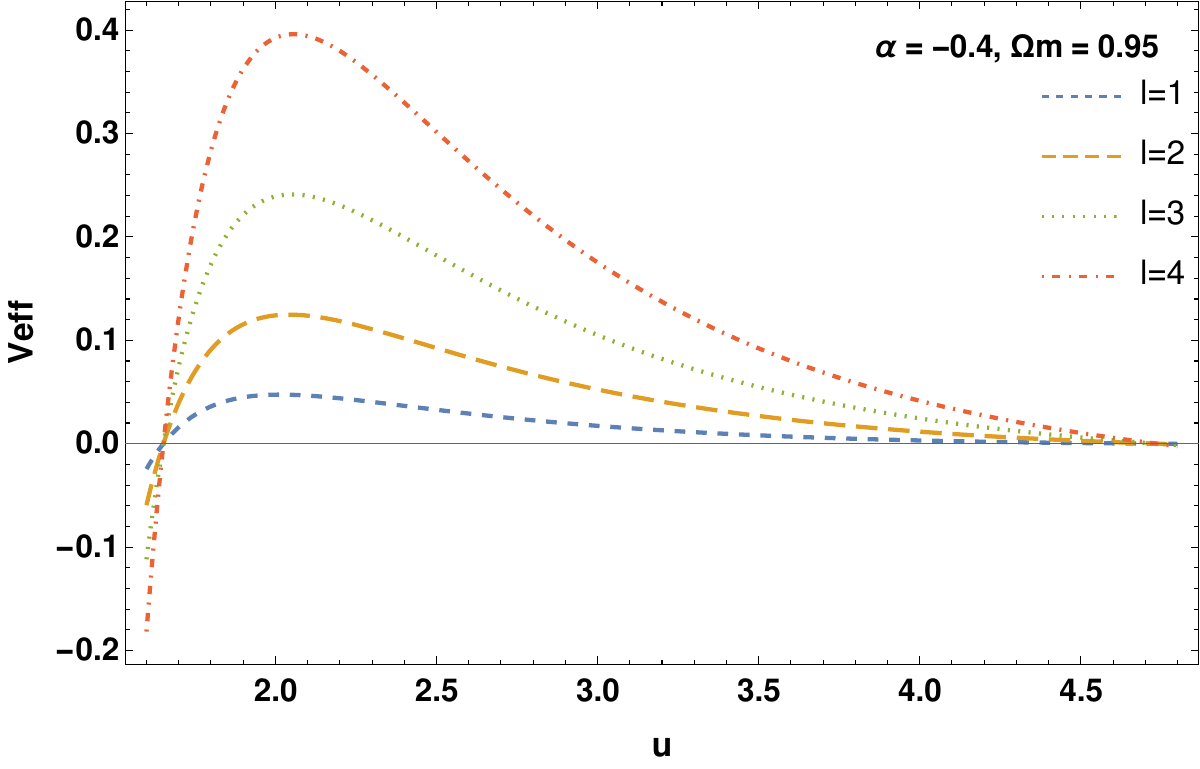}
	\endminipage
	\caption{Effective potentials ($V_\text{eff}$) for scalar perturbations as functions of radial distance $u$. Each panel corresponds to a different value of the MGCG parameter $\alpha$ (increasing from left to right), with $\Omega_m = 0.95$ fixed. Colored curves represent different multipole numbers $l$. The leftmost plot corresponds to the near-extremal black hole configuration.}
\label{fig_potentialo95}
\end{figure*}
\begin{figure}
	\centering
	\includegraphics[width=0.85\linewidth]{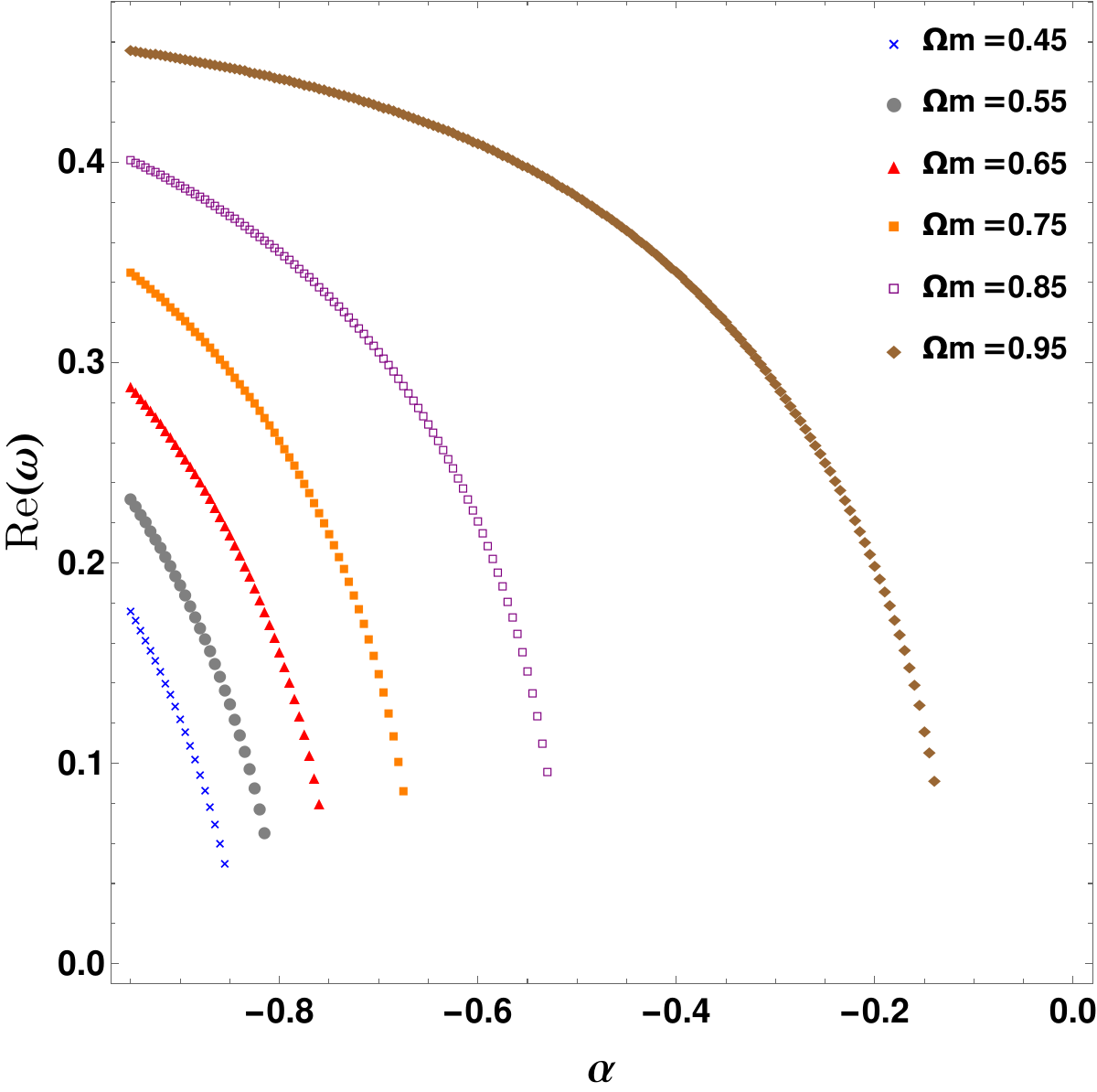}
	\caption{Variation of the real part of the QNM frequency with the MGCG parameter $\alpha$ for scalar field perturbations. The plot shows results for the fundamental mode with multipole number $l = 2$. Different colored curves correspond to different values of the cosmological parameter $\Omega_m$.} \label{fig:frequencyplotvsalpha}
\end{figure}
There is a need to solve the second-order differential equation. Various methods are available, each with its own advantages and limitations \cite{berti_qnm_review}. Among them, the WKB approximation-an analytical technique\cite{Iyer:1986np}-is widely used to determine quasinormal mode frequencies. Given the complexity of the metric, the WKB method offers a more tractable approach. 
\begin{figure*}[htb!]
	\centering
	\minipage{0.33\textwidth}
	\includegraphics[width=\linewidth]{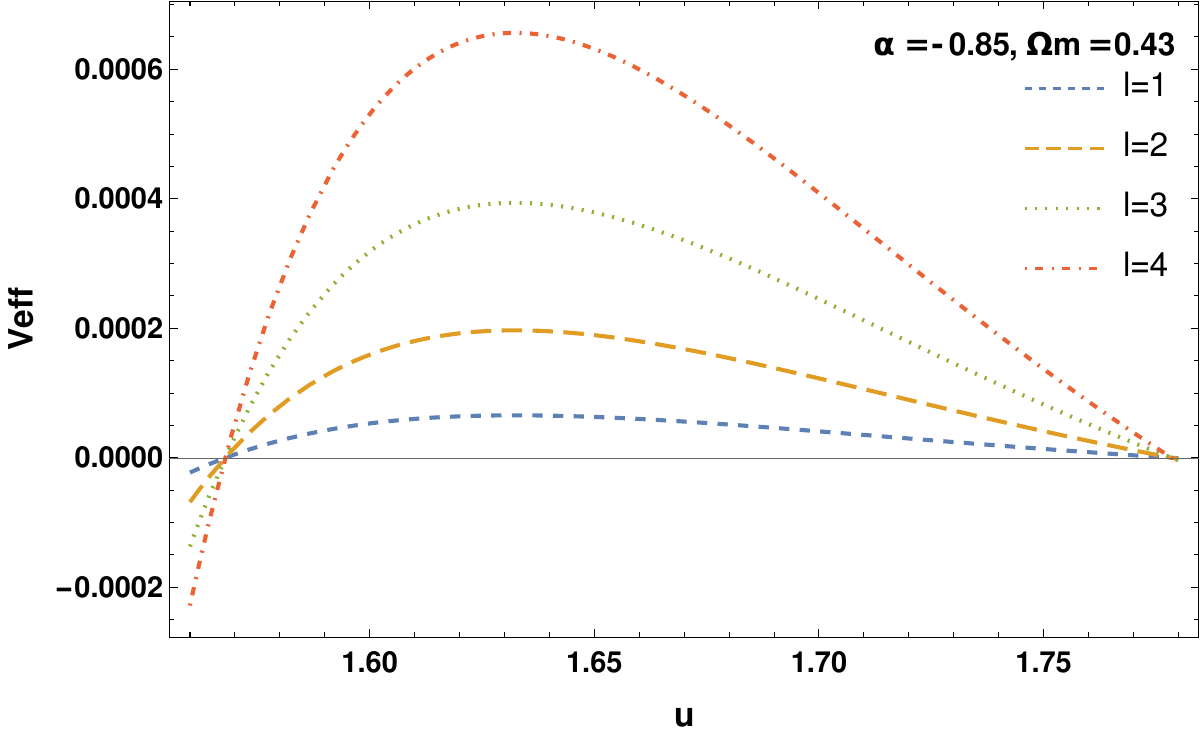}
	\endminipage\hfill
	\minipage{0.33\textwidth}
	\includegraphics[width=\linewidth]{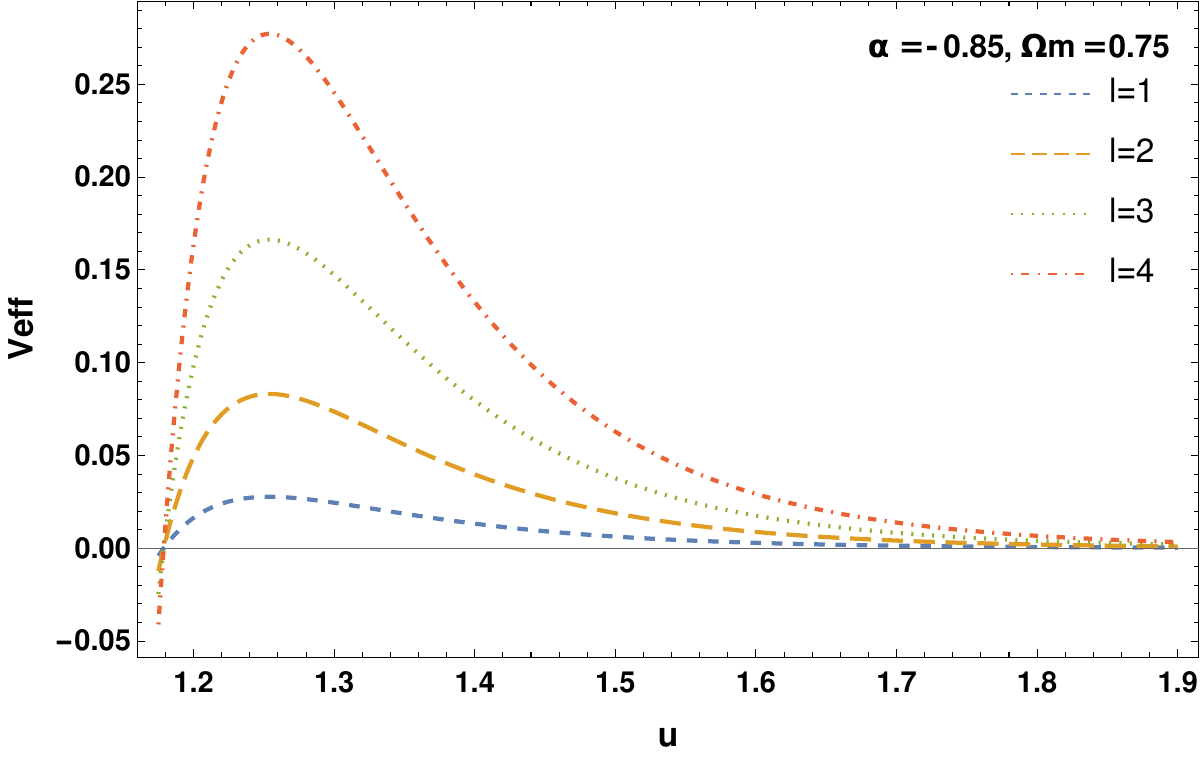}
	\endminipage
 \hfill
    \minipage{0.33\textwidth}
	\includegraphics[width=\linewidth]{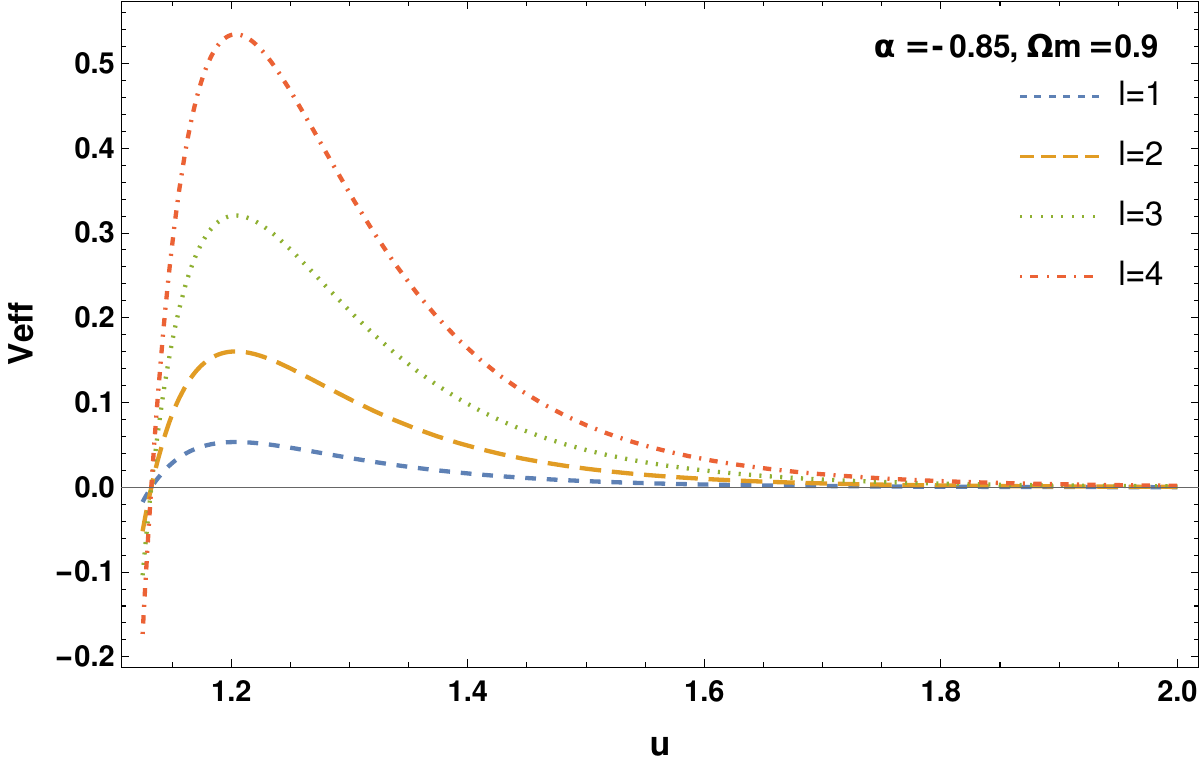}
	\endminipage
\caption{Effective potentials ($V_\text{eff}$) for electromagnetic perturbations as functions of radial distance $u$. Plots are shown for different multipole numbers $l$, with the MGCG parameter fixed at $\alpha = -0.85$ and varying values of $\Omega_m$. The colored curves illustrate how the potential structure depends on $l$}
\label{fig_potentiala85e}
\end{figure*}	

\subsection{Electromagnetic field}
A massless vector field is influenced by the curvature of the blackhole, which actively governs its dynamics. The evolution of the field is described by Maxwell’s equations in  curved spacetime. \\Due to its static and spherically symmetric nature, the spacetime associated with the MCGC blackhole admits a simplified geometry. The complete 4-dimensional manifold can be decomposed as $\mathbb{R}^2 \times S^2$, separating the temporal-radial and angular parts of the spacetime. This decomposition allows for a more \textbf{systematic} analysis of field dynamics \cite{wald_gr}.
\begin{equation}
\nabla_\mu F^{\mu\nu} = 0
\end{equation}
In this background, the behavior of a vector field can be studied through parity decomposition using spherical harmonics defined on the $S^2$ sphere. The field separates into two distinct sectors: odd (axial) and even (polar) parity components \cite{moncrief1974odd}.

Because the Maxwell equations are linear and the spacetime possesses spherical symmetry, each parity mode evolves independently at linear order in perturbation theory-that is, there is no coupling between them \cite{chandrasekhar1985mathematical}. This allows each parity component to be treated separately when analyzing the dynamics of the vector field .

Substituting the parity-decomposed vector field into the Maxwell equations leads to a radial differential equation. This equation takes the form of a Schrödinger-like wave equation  \cite{Rahman:2023swz}, which governs the evolution of each parity mode.
\begin{figure}
	\centering
	\includegraphics[width=0.85\linewidth]{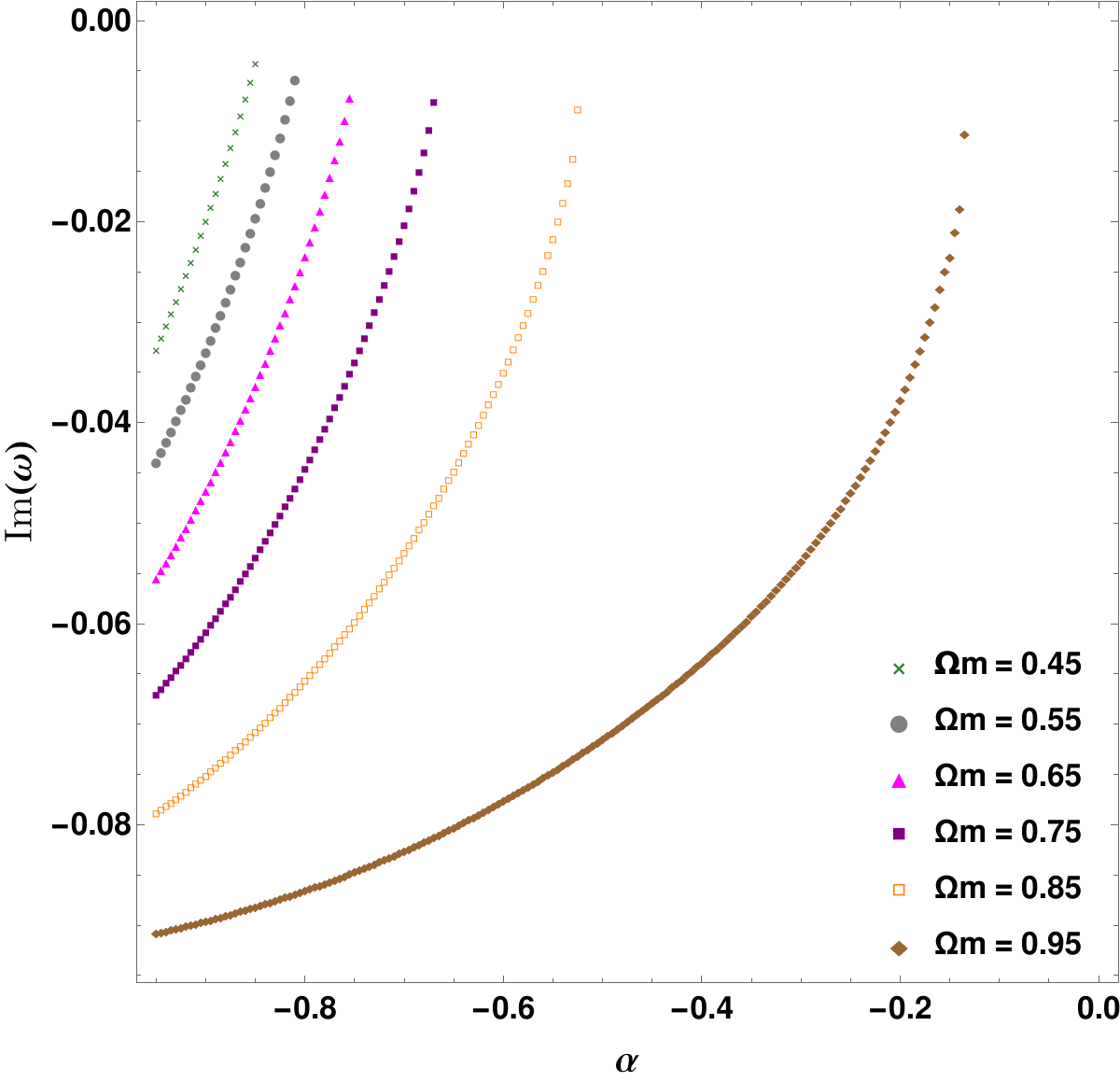}
	\caption{Imaginary part of the QNM frequency for scalar field perturbations as a function of the MGCG parameter $\alpha$. The plot corresponds to the fundamental mode with $l = 2$, with different colored curves representing various values of the cosmological parameter $\Omega_m$.}
	\label{fig:decayplotvsalpha}
\end{figure}
\begin{figure}
	\centering
	\includegraphics[width=0.85\linewidth]{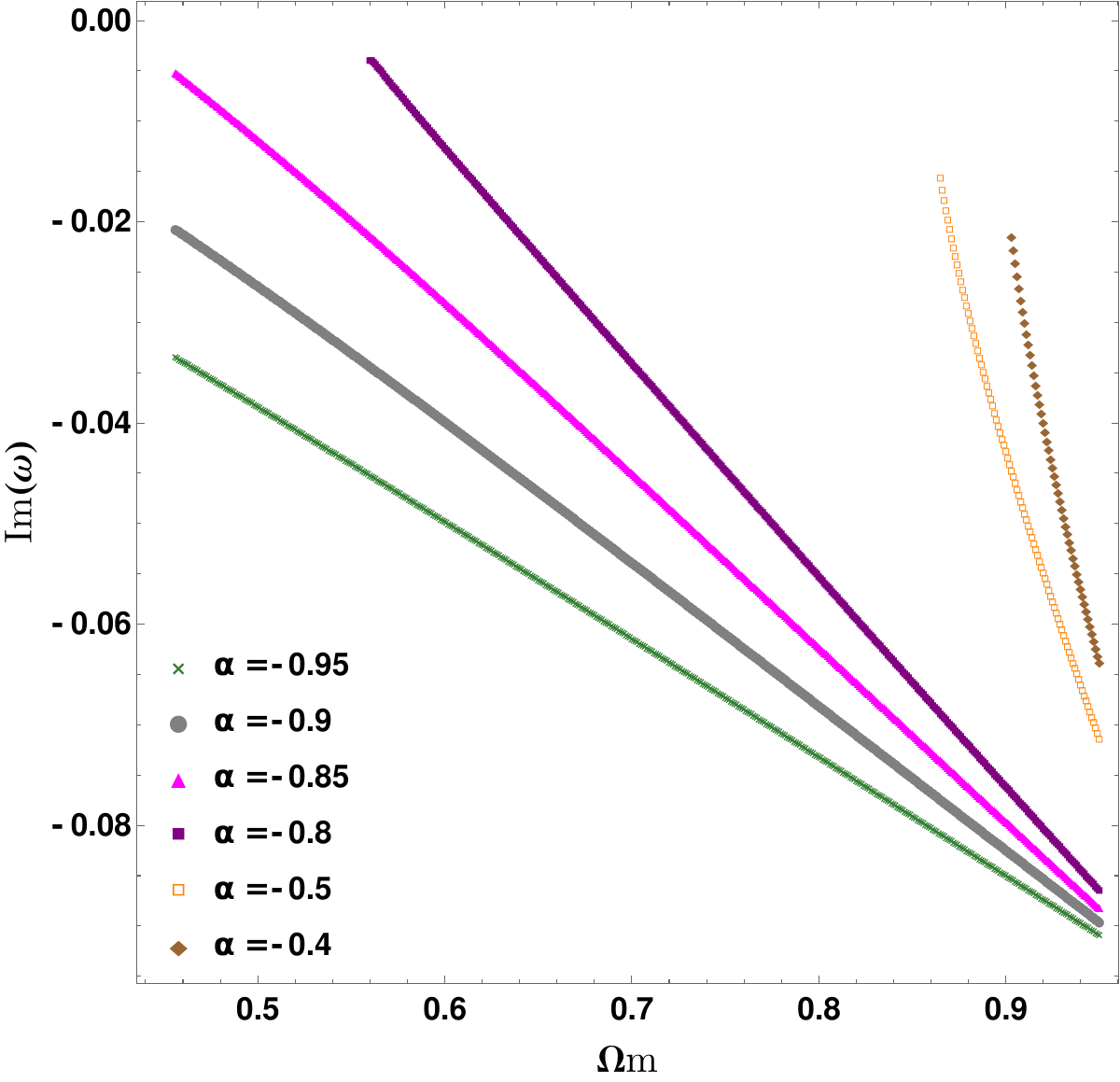}
	\caption{Imaginary part of the QNM frequency as a function of $\Omega_m$ for scalar field perturbations. The plot corresponds to the fundamental mode with $l = 2$. Different colored curves represent various values of the MGCG parameter $\alpha$.}
	\label{fig:decayplotvsomega}
\end{figure}
Interestingly, in this MCGC background - as in other spherically symmetric spacetimes like Schwarzschild -the resulting Schrödinger-type equations for both odd and even parity modes feature the same effective potential \cite{price1972nonspherical}.

  \begin{equation}
 V^{el}_{eff}(u)= f(u) \cdot \left( \ell (\ell + 1) \, u^{-\frac{2}{\alpha + 1}} \right)
\end{equation}
 \begin{figure*}[htb!]
	\centering
	\minipage{0.49\textwidth}
	\includegraphics[width=\linewidth]{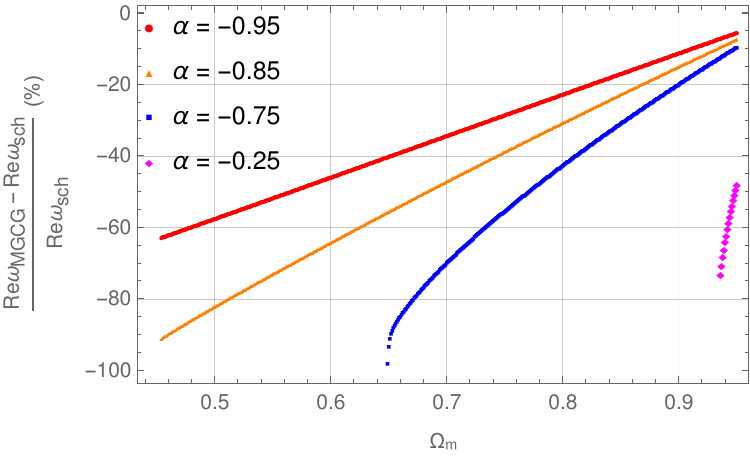}
	\endminipage
    \hfill
	\minipage{0.49\textwidth}
	\includegraphics[width=\linewidth]{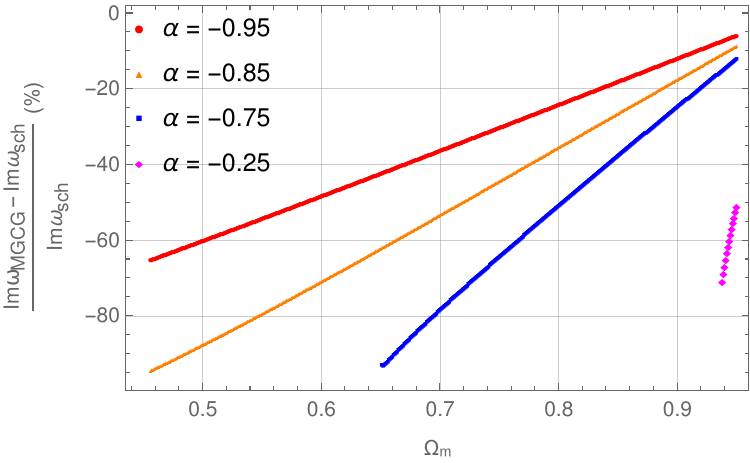}
	\endminipage
    \hfill
\caption{Relative deviations in QNM frequencies of the MGCG black hole from Schwarzschild as a function of $\Omega_m$. The left panel shows the normalized difference in the real part: $(\text{Re}[\omega_{\text{MGCG}}] - \text{Re}[\omega_{\text{Sch}}])/\text{Re}[\omega_{\text{Sch}}]$, while the right panel presents the corresponding difference in the imaginary part. Each curve corresponds to a different value of the model parameter $\alpha$, as indicated by color. The QNM data correspond to scalar perturbations, with all results shown for the fundamental mode with $l = 2$.}
 \label{fig_qnm_diff_l2}
\end{figure*}	
Unlike the scalar field case, the effective potential for the vector field depends only on the horizon function-not on its derivative. 
 \begin{figure*}[htb!]
	\centering
	\minipage{0.49\textwidth}
	\includegraphics[width=\linewidth]{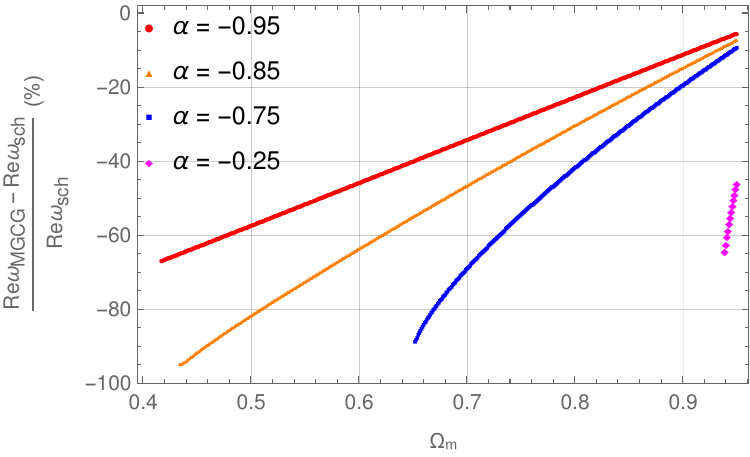}
	\endminipage
    \hfill
	\minipage{0.49\textwidth}
	\includegraphics[width=\linewidth]{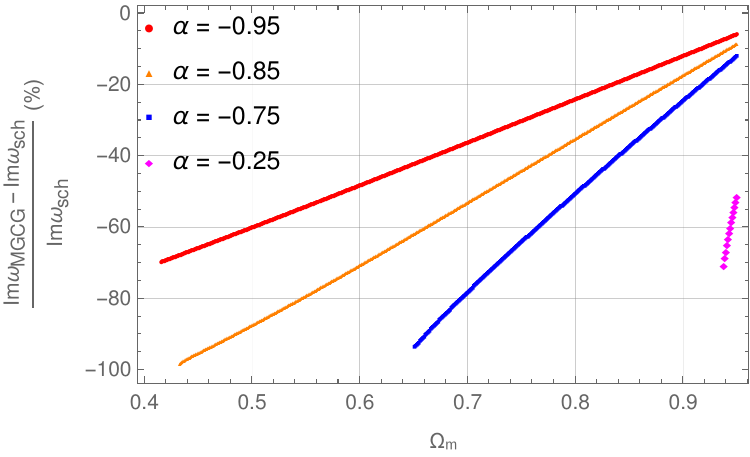}
	\endminipage
    \hfill
\caption{Relative Quasi-Normal modes deviations of the MGCG blackhole from Schwarzschild, as functions of $\Omega_m$. The left panel shows the real part, and the right panel shows the imaginary part of the normalized frequency differences. Shown are results for electromagnetic perturbations, with each coloured curve corresponding to a different $\alpha$. All results are for the $l = 2$ fundamental mode.} \label{fig_qnm_diff_em_l2}
\end{figure*}	

  \section{Effective Potential}\label{effective pot}

  \subsection{Effective Potential of Scalar Perturbation} The structure of the effective potential is too complex to handle exactly in closed analytical form. To manage this, we expand the potential around a suitable point, so that at a certain order of expansion, it captures the behavior of the potential within the required radial range. The effective potential for scalar perturbations in the MCGM blackhole spacetime shows that the horizon structure of the blackhole is independent of the perturbation or the nature of the perturbing field. It depends solely on the blackhole parameters-in this case, the $\Omega_{m}$ and $\alpha$ parameters.
However, the peak of the potential-corresponding to the photon orbit \cite{Cardoso2009} is influenced by both the blackhole parameters and the perturbing field. Plots of the effective potential offer useful guidance in dealing with the challenges of locating the maximum of the potential, which is essential for applying the WKB method.

One important observation in Fig.~(\ref{fig_potentialo95}) is that the peak of the potential usually lies close to the event horizon. To estimate its position, we first expand the potential near this region and compute an approximate value of the maximum. Then, using the \texttt{FindRoot} function in \textit{Mathematica}, the result can be refined by providing a close initial guess.

Another observation is that the region of interest-bounded by the event horizon and cosmological horizon-shrinks as the value of $\alpha$ becomes more negative, with $\Omega_{m}$ held fixed. The two horizons approach each other but do not overlap, preserving the blackhole’s horizon structure. Similarly, decreasing the matter energy density compresses the space between the two horizons. These features are clearly reflected in the graphs of the effective potential.
\begin{figure*}[htb!]
	\centering
	\minipage{0.33\textwidth}
	\includegraphics[width=\linewidth]{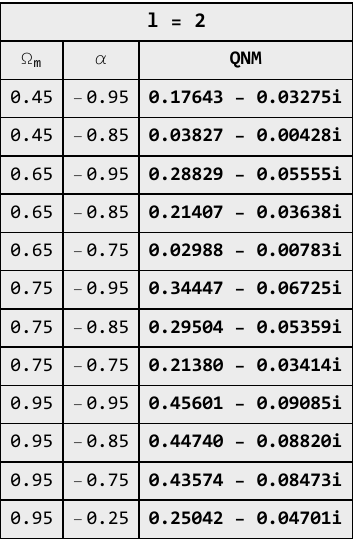}
	\endminipage 
    \hfill
	\minipage{0.33\textwidth}
	\includegraphics[width=\linewidth]{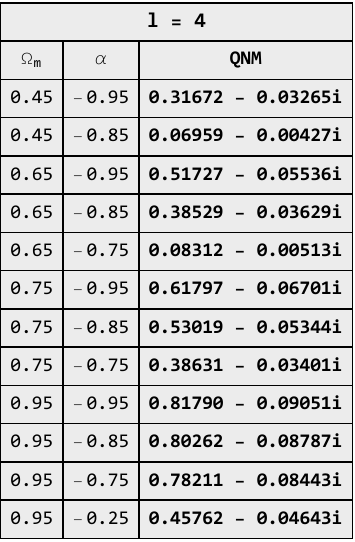}
	\endminipage
 \hfill
	\minipage{0.33\textwidth}
	\includegraphics[width=\linewidth]{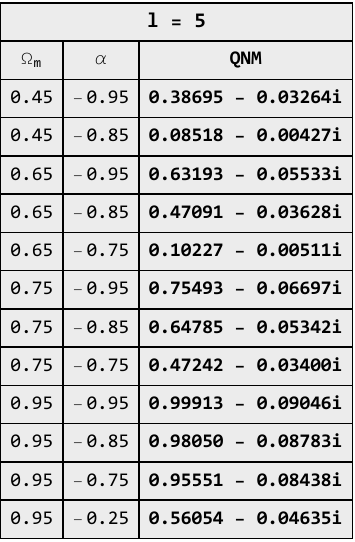}
	\endminipage
	\caption{Fundamental(\(n = 0\)) quasi-normal mode frequencies for scalar perturbations in the modified Chaplygin gas blackhole background for multipole indices \(l = 2, 4, 5\).}\label{QNMfundatables}
\end{figure*}
As the region between the horizons shrinks, the maximum value of the potential also decreases. This maximum significantly influences the oscillation frequency of the perturbations Fig.~(\ref{fig:frequencyplotvsalpha}), while the sharpness of the peak governs the decay rate. As the multipole number $l$ increases, both the height and sharpness of the potential peak increase, leading to higher oscillation frequencies and faster decay of the perturbations.

\subsection{Effective Potential for Electromagnetic Perturbation}
A comparison between the scalar and electromagnetic perturbations shows that their effective potentials differ across the spacetime, with this difference ranging typically from the order of $10^{-6}$ to $10^{-2}$. In the region of Fig.~(\ref{fig_potentiala85e}) where the event horizon and the cosmological horizon are very close to each other, the difference becomes quite small-about $10^{-6}$. Near the event horizon, the effective potential for scalar field perturbations is generally larger, while near the cosmological horizon, the effective potential for electromagnetic perturbations is greater. This behavior reflects the distinct nature of the two fields and how they interact differently with the geometry of spacetime, rather than indicating that one perturbation is dominant over the other.

As the parameter $\alpha$ increases, the distance between the horizons also increases. This leads to a more noticeable difference in the effective potentials, especially near the event horizon, where the gap can reach up to the order of $10^{-2}$. In contrast, this difference becomes increasingly small near the cosmological horizon. These trends suggest that the effective potential is sensitive not only to the type of perturbing field but also to the geometry of the blackhole background, as governed by the parameters of the MGCG model.	
 \section{WKB Method}\label{WKB}
 The WKB (Wentzel--Kramers--Brillouin)\cite{Schutz:1985km} method is a semi-analytic technique for approximating solutions to second-order differential equations, particularly in quantum mechanics and black hole perturbation theory. Several approaches exist within the WKB regime, but the core of the method lies in imposing quantization conditions on wave-like solutions across potential barriers.
\begin{figure*}[htb!]
	\centering
	\minipage{0.33\textwidth}
	\includegraphics[width=\linewidth]{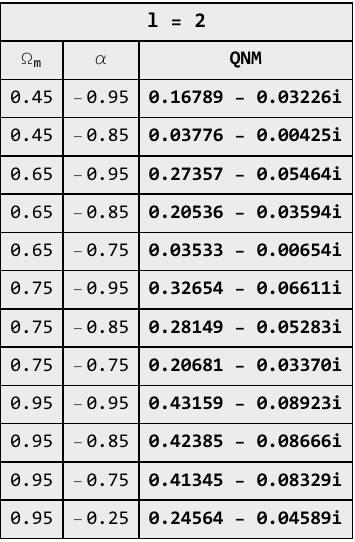}
	\endminipage\hfill
	\minipage{0.33\textwidth}
	\includegraphics[width=\linewidth]{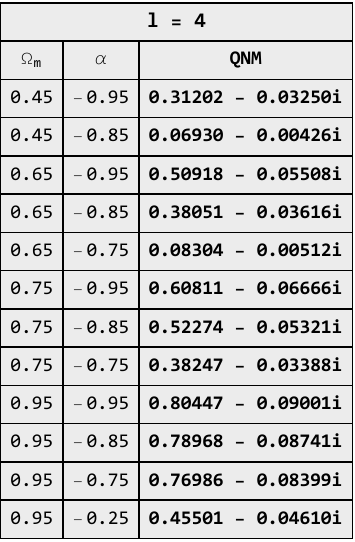}
	\endminipage
 \hfill
	\minipage{0.33\textwidth}
	\includegraphics[width=\linewidth]{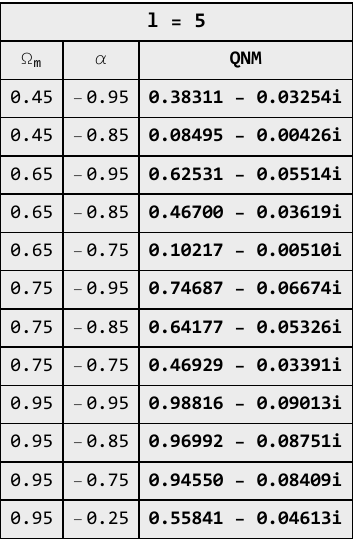}
	\endminipage
	\caption{Fundamental (\(n = 0\)) quasi-normal mode frequencies for electromagnetic perturbation in the modified Chaplygin gas blackhole background for multipole indices \(l = 2, 4, 5\).}\label{QNMn=funda_table_el}
\end{figure*}	
Iyer and Will \cite{Iyer:1986nq} extended the standard WKB approach to third order and applied it to compute quasi-normal modes (QNMs) of blackholes. Near the peak of the effective potential, they performed a Taylor expansion up to the required order, then derived an asymptotic approximation to the interior solution. This was matched with WKB solutions from the exterior regions to determine the connection coefficients.
 \begin{figure*}[htb!]
	\centering
	\minipage{0.49\textwidth}
	\includegraphics[width=\linewidth]{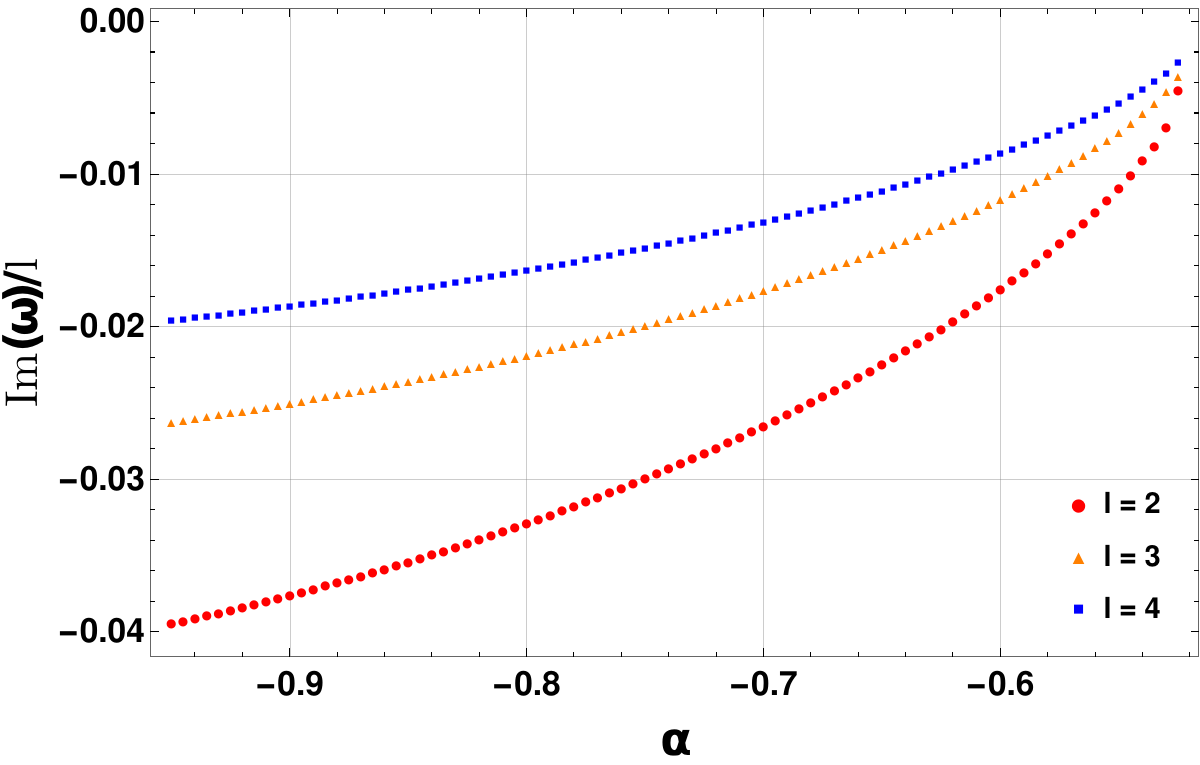}
	\endminipage
    \hfill
	\minipage{0.49\textwidth}
	\includegraphics[width=\linewidth]{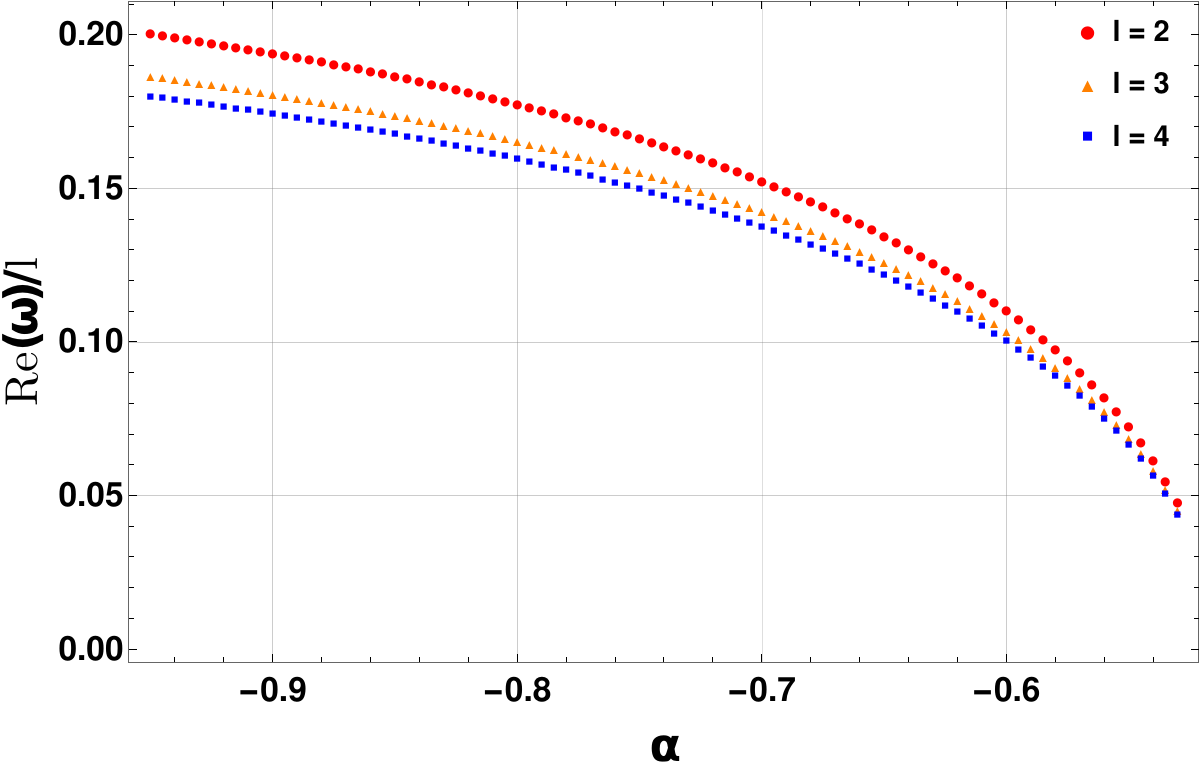}
	\endminipage
    \hfill
\caption{Normalized QNM frequencies as functions of the MGCG parameter $\alpha$. The left panel shows $\text{Im}(\omega)/l$, and the right panel shows $\text{Re}(\omega)/l$, plotted for scalar perturbations. Different colors and markers represent QNM data for various multipole numbers $l$. Deviations from linearity reflect the influence of the MGCG model.}
\label{normalisedQNMplot}
\end{figure*}
In our case, the effective potential governing scalar and electromagnetic perturbations in the MGCG model satisfies the WKB applicability conditions\cite{Konoplya:2019hlu}. We have further extended the method to sixth order. Our correction terms at each WKB order show strong agreement with the results of Konoplya \cite{Konoplya:2003ii}.
Regarding boundary conditions: the event horizon behaves like a one-way membrane, waves can enter but not escape. Similarly, at the cosmological horizon, only outgoing waves are permitted, with no incoming radiation from that boundary. These boundary conditions lead to a discrete set of quasi-normal frequencies.
 \begin{equation}
           \frac{d^2 \Psi(u_*)}{du_*^2}  + Q(u)=0
       \end{equation}
    Where $$Q(u)= \omega^2 - V_{\text{eff}}(u) $$
     Expand the $Q(u)$ in a Taylor  Series about the point $u_{0}$, at which $Q(u)$ reaches a minimum (or $V(u)$ is maximum)
\begin{multline}
Q(u) = Q_0 + \frac{Q_0''}{2!}(u - u_0)^2 + \frac{Q_0^{(3)}}{3!}(u - u_0)^3 
+ \frac{Q_0^{(4)}}{4!}(u - u_0)^4  \\ + \frac{Q_0^{(5)}}{5!}(u - u_0)^5 
+ \frac{Q_0^{(6)}}{6!}(u - u_0)^6 + \frac{Q_0^{(7)}}{7!}(u - u_0)^7 
+ \frac{Q_0^{(8)}}{8!}(u - u_0)^8 \\ + \frac{Q_0^{(9)}}{9!}(u - u_0)^9 
+ \frac{Q_0^{(10)}}{10!}(u - u_0)^{10} + \frac{Q_0^{(11)}}{11!}(u - u_0)^{11} \\
+ \frac{Q_0^{(12)}}{12!}(u - u_0)^{12}
\end{multline}

For the computation of the $6^{\text{th}}$-order WKB frequency, the Taylor expansion of the effective potential must include terms up to at least the $12^{\text{th}}$ derivative.

\begin{equation}\label{WKB eq}
    \omega^{2}= n+\frac{1}{2} + V_{eff}(u_{0}) + \epsilon^{2} \Lambda_{2}+\epsilon^{3} \Lambda_{3} + \epsilon^{4} \Lambda_{4} + \epsilon^{5} \Lambda_{5} + \epsilon^{6} \Lambda_{6} 
    \end{equation}
Here, $\epsilon$ represents the order of the perturbation, and $\Lambda_{i}$ denotes the $i^{\text{th}}$-order correction to the quasi-normal mode frequency.The corresponding expressions are too lengthy to include in the paper but can be provided upon request if needed. By putting all the $\Lambda_{i}$ terms into Eq.(\ref{WKB eq}), we can determine the quasi-normal mode frequency.
\begin{equation}\label{QNM exp}
    \omega^{2}=  V_{eff}(u_{0}) +\sqrt{-2 V^{(2)}_{u_0}}\biggl( \mathrm{i}\,\beta\left(1 +\Lambda_{3} + \Lambda_{5}\right) + \left(\Lambda_{2} -  \Lambda_{4} +  \Lambda_{6}\right)\biggr)
\end{equation}
   \section{Results} \label{result}

The imaginary part of the quasi-normal mode (QNM) frequency $\omega$ encodes information about the stability of blackhole perturbations. Although it depends on both the model parameters and the nature of the perturbing field, the qualitative diagnosis of stability is largely independent of the specific type of perturbation. In Fig.(\ref{fig:decayplotvsalpha}), the imaginary part of $\omega$ is analyzed as a function of the Chaplygin gas parameter $\alpha$ for several fixed values of the matter energy density $\Omega_{m}$.
	
For a fixed \(\Omega_{m}\), we observe that the decay rate of the perturbations increases as \(\alpha\) becomes more negative. This indicates that perturbations dissipate more rapidly in the regime of strongly negative \(\alpha\). This trend is consistent across all considered values of \(\Omega_{m}\). Additionally, the slope of the increase in decay rate is steeper for less negative \(\alpha\), i.e., when the event horizon and cosmological horizon are in closer proximity. As the separation between the horizons increases, this rate of change becomes more gradual. This behavior is more pronounced for higher values of \(\Omega_{m}\).

Fig.(\ref{fig:decayplotvsomega}) reveals the  effect of matter energy density \(\Omega_{m}\) on the stability of the blackhole is also examined. It is found that the decay rate increases with increasing \(\Omega_{m}\), implying that black holes become more stable as the parameter of matter energy density grows. For a fixed value of \(\alpha\), the slope at which the decay rate increases with \(\Omega_{m}\) remains nearly constant along the curve, although the overall tendency to increase is more prominent for less negative values of \(\alpha\).
Fig.(\ref{fig_qnm_diff_l2}) and Fig.(\ref{fig_qnm_diff_em_l2}) reveal that the quasinormal mode (QNM) frequencies in the MGCG blackhole model deviate noticeably from those of the Schwarzschild blackhole, with the degree of deviation depending sensitively on both the Chaplygin gas parameter $\alpha$ and the matter density $\Omega_m$. These deviations are substantial - well beyond what typical observational uncertainties could conceal.
\begin{figure*}[htb!]
	\centering
	\minipage{0.33\textwidth}
	\includegraphics[width=\linewidth]{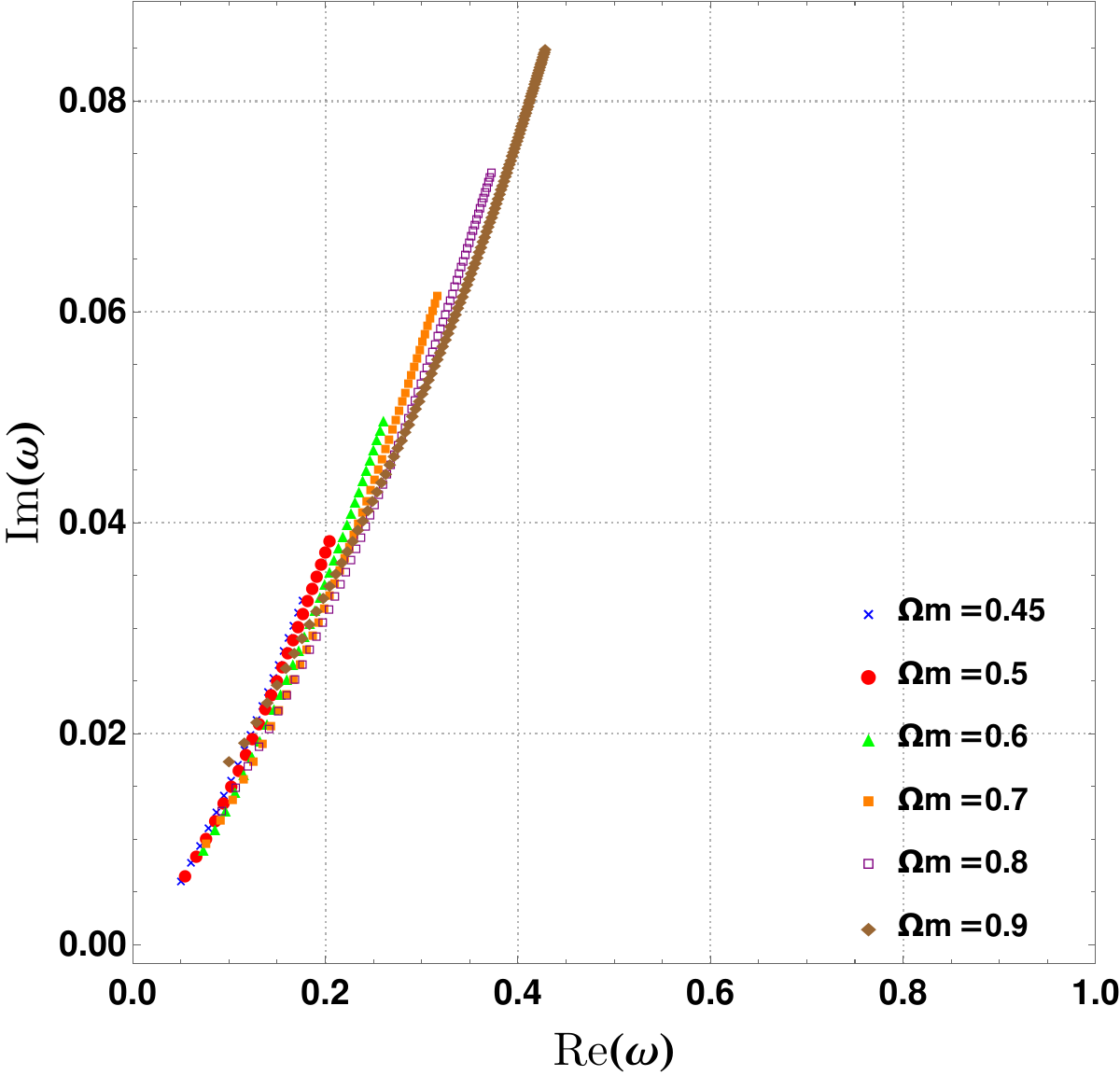}
	\endminipage\hfill
	\minipage{0.33\textwidth}
	\includegraphics[width=\linewidth]{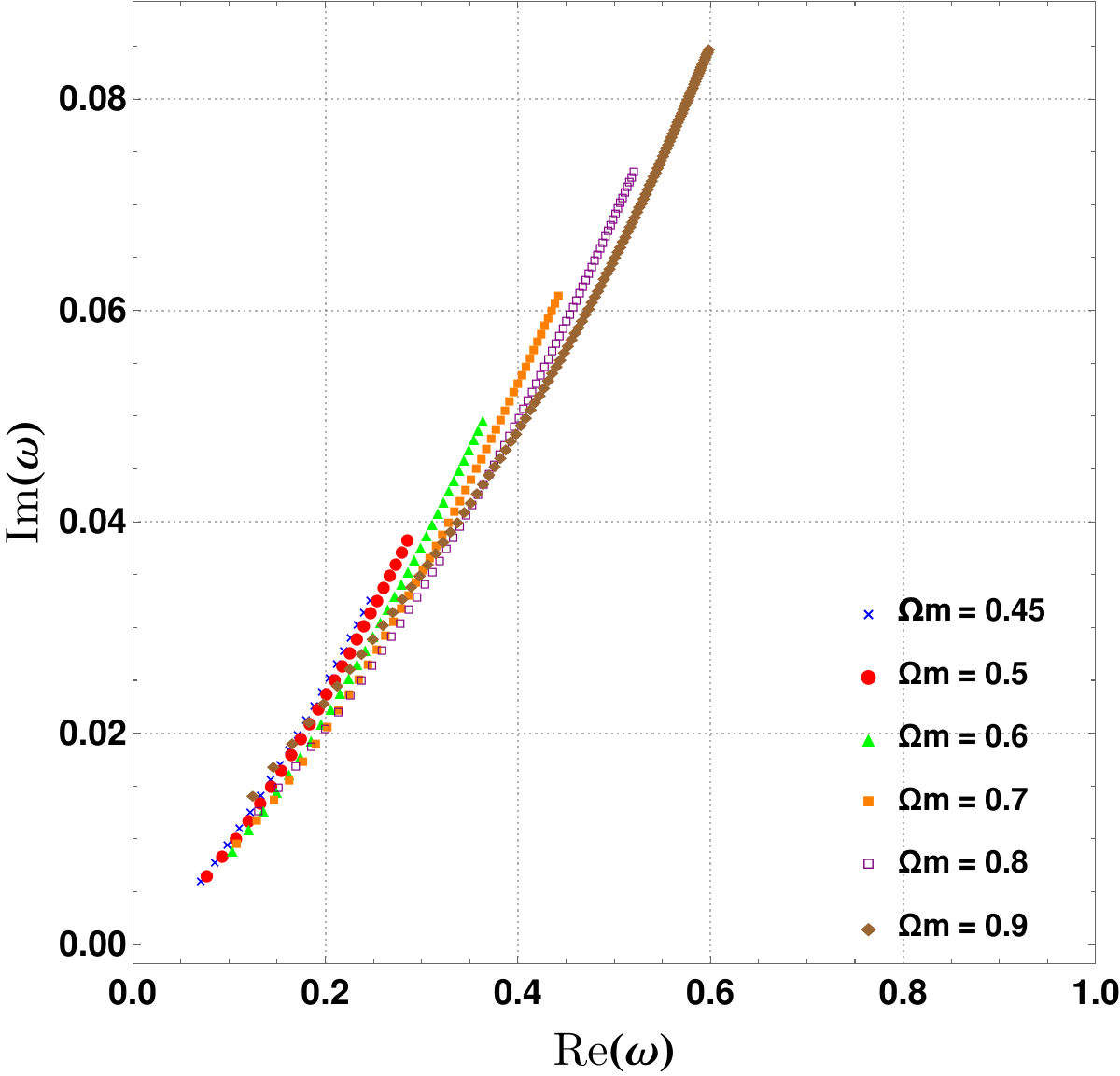}
	\endminipage
 \hfill
	\minipage{0.33\textwidth}
	\includegraphics[width=\linewidth]{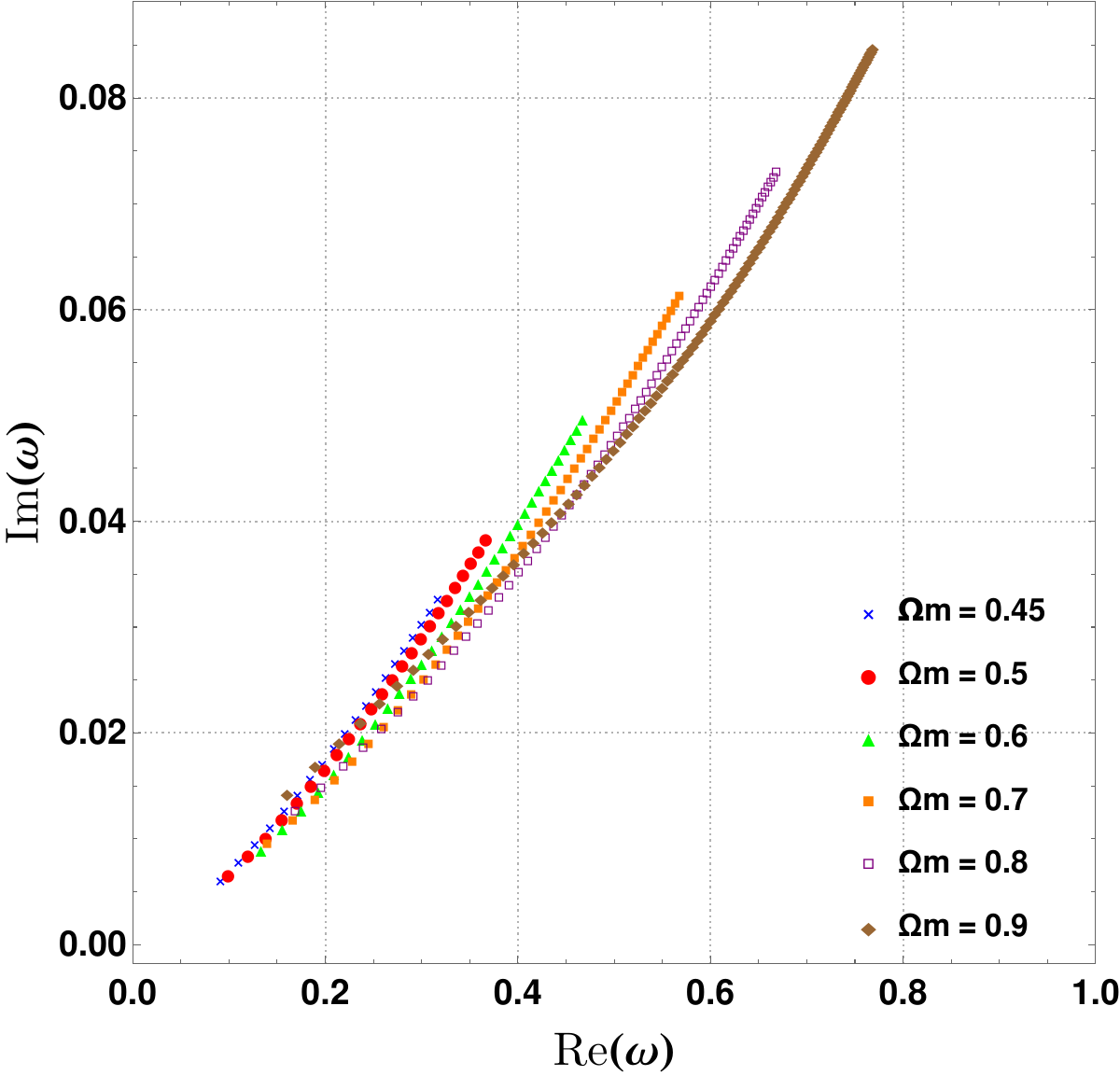}
	\endminipage
	\caption{Curves of fundamental QNM frequencies in the complex conjugate plane for scalar field perturbations. Panels from left to right correspond to $l = 2$, 3, and 4. Each colored curve represents a fixed $\Omega_m$, with points along the curve showing the variation as the MGCG parameter $\alpha$ changes.}
 \label{complexplots}
\end{figure*}	
The relative deviation varies with $\alpha$, reaching a peak at intermediate values, while both more negative and less negative values lead to smaller deviations.
\begin{figure*}[htb!]
	\centering
	\minipage{0.33\textwidth}
	\includegraphics[width=\linewidth]{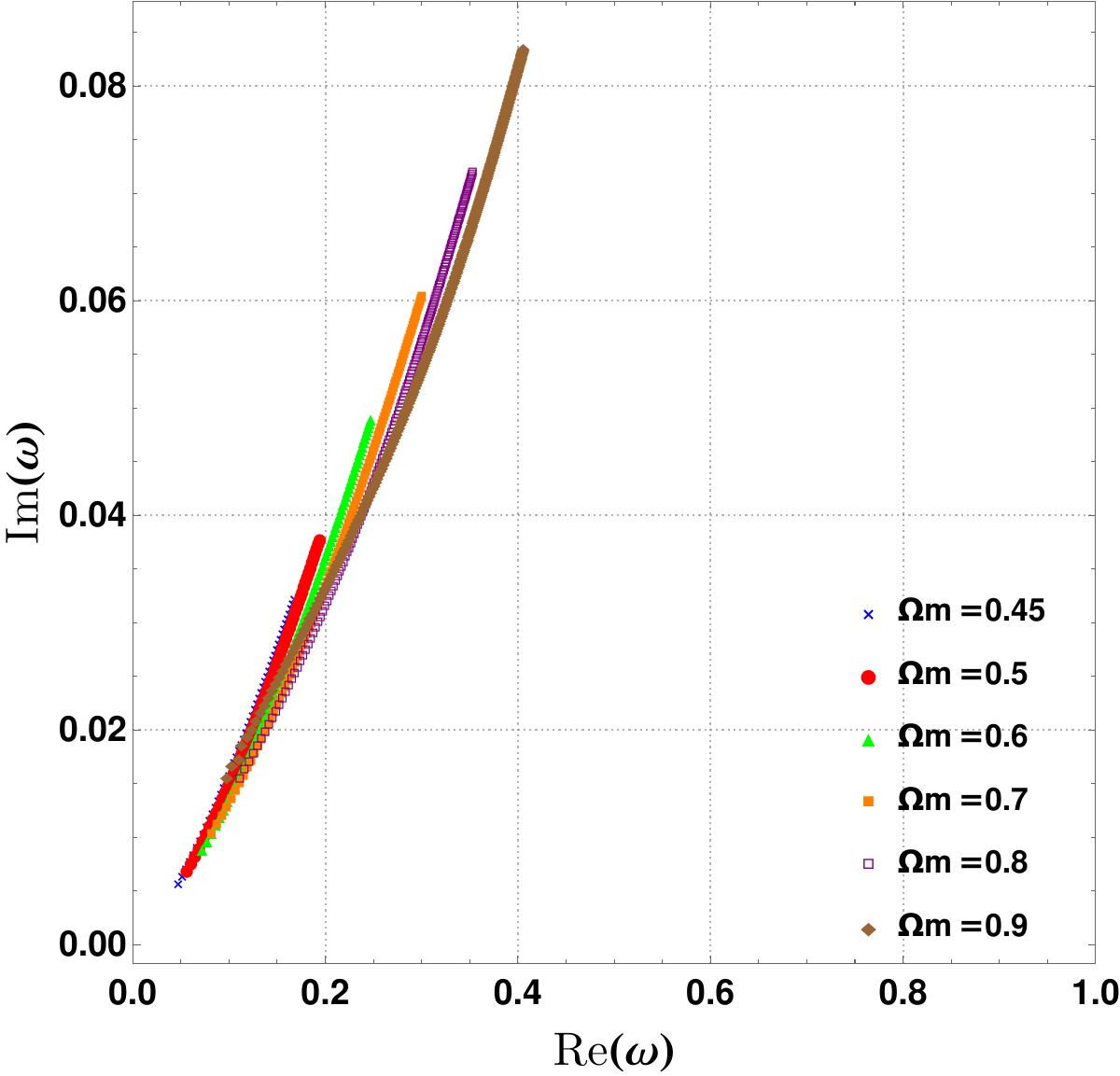}
	\endminipage\hfill
	\minipage{0.33\textwidth}
	\includegraphics[width=\linewidth]{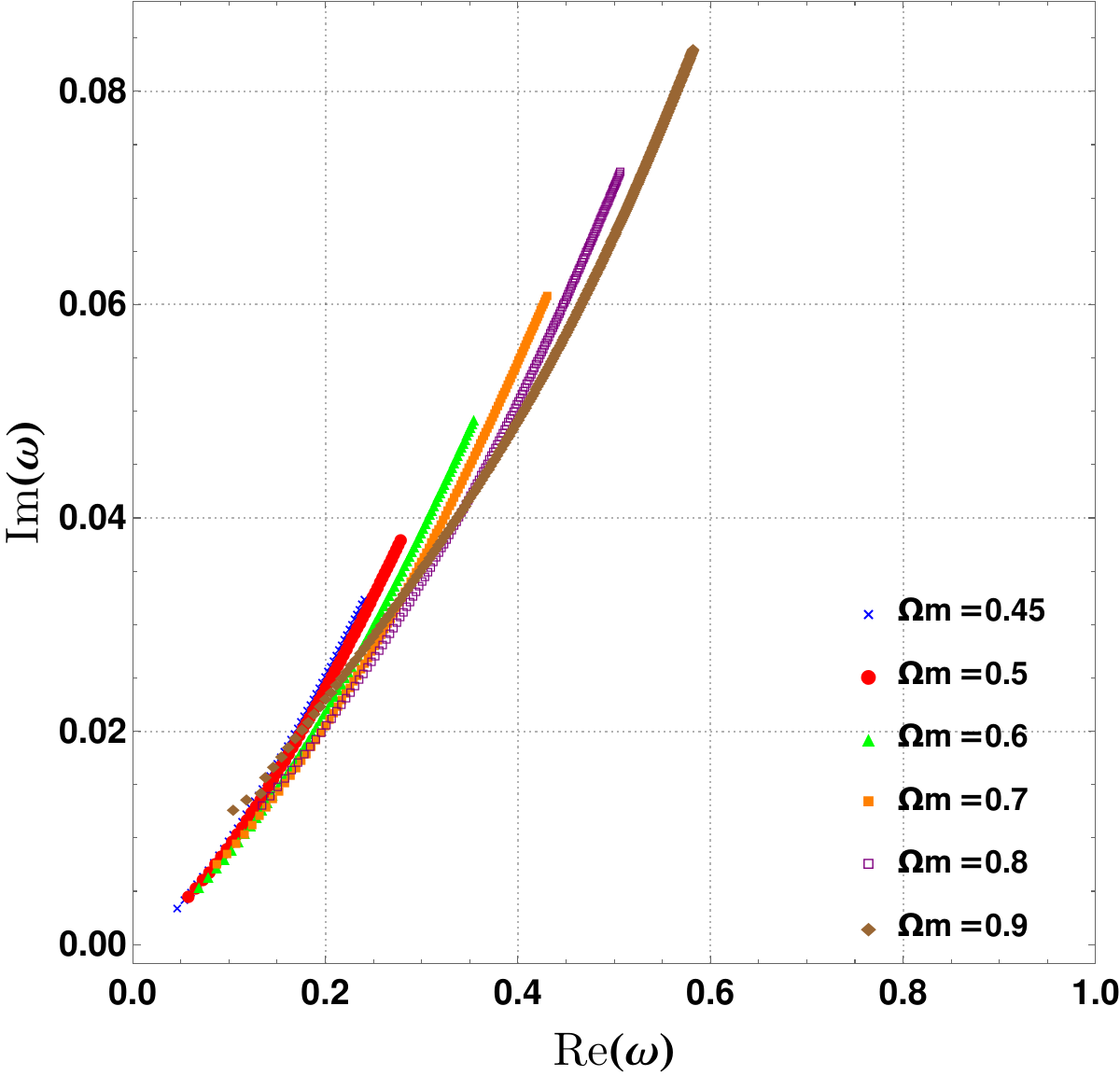}
	\endminipage
 \hfill
	\minipage{0.33\textwidth}
	\includegraphics[width=\linewidth]{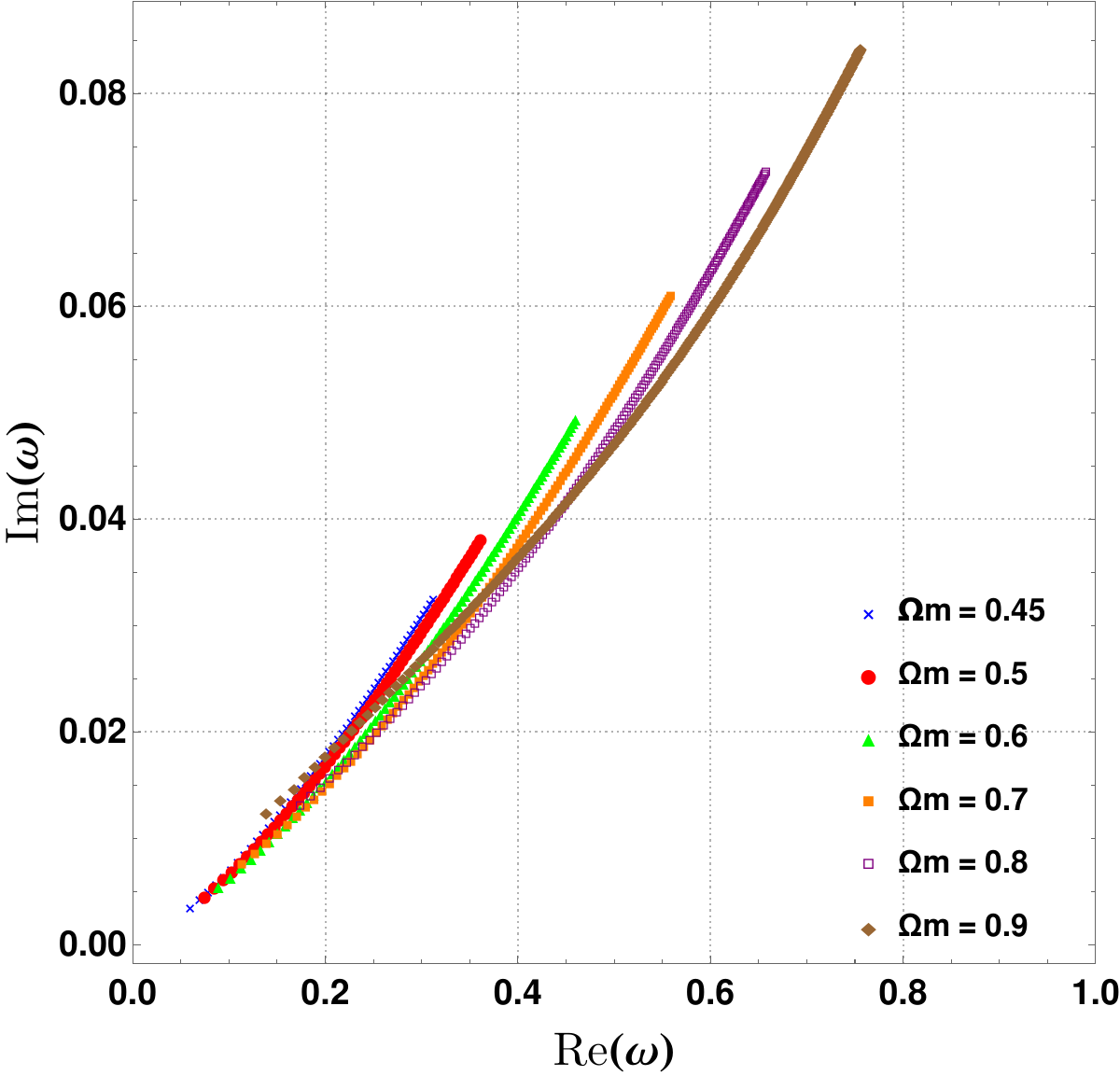}
	\endminipage
	\caption{Fundamental QNM frequencies in the complex conjugate plane for electromagnetic perturbations. Panels (left to right) correspond to $l = 2$, 3, and 4. Colored curves represent fixed $\Omega_m$, with variation along each curve due to changing $\alpha$.}
 \label{complexplotem}
\end{figure*}

The matter density parameter $\Omega_m$ also plays a suppressive role. As $\Omega_m$ increases, the QNM frequencies increasingly resemble those of the Schwarzschild case. This trend is consistently observed in both the real and imaginary components of the spectrum and holds across the full range of $\alpha$.
The table (\ref{QNMfundatables}) and (\ref{QNMn=funda_table_el}) shows the variation of QNM frequencies with the multipole number. Differences begin to arise at the fourth order.  however, these differences are not visible in the plots. 
\begin{figure*}[htb!]
	\centering
	\minipage{0.33\textwidth}
	\includegraphics[width=\linewidth]{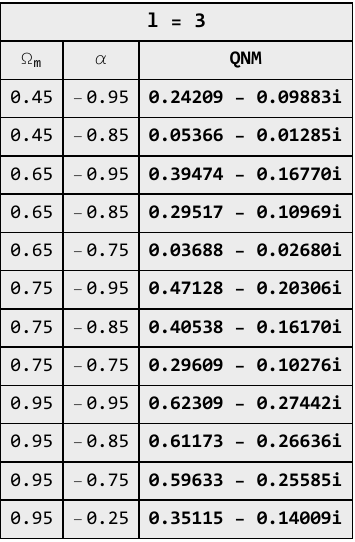}
	\endminipage\hfill
	\minipage{0.33\textwidth}
	\includegraphics[width=\linewidth]{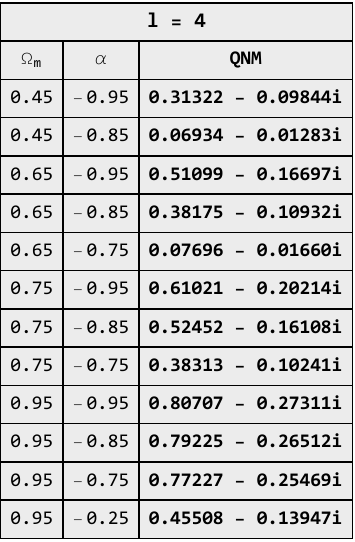}
	\endminipage
 \hfill
	\minipage{0.33\textwidth}
	\includegraphics[width=\linewidth]{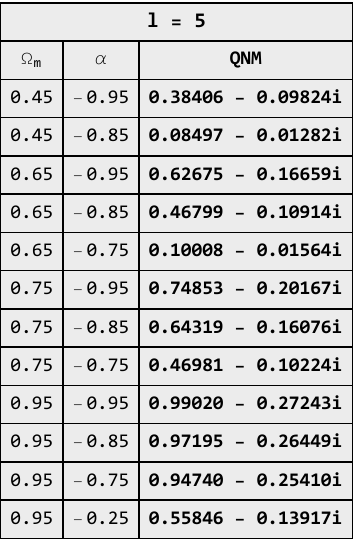}
	\endminipage
	\caption{First overtone (\(n = 1\)) quasi-normal mode frequencies for scalar perturbations in the modified Chaplygin gas blackhole background for multipole indices \(l = 3, 4, 5\).}\label{QNMn=1tables}
\end{figure*}	
Importantly, both the real and imaginary parts of the QNM frequencies exhibit significant relative deviations from the Schwarzschild case across most of the parameter space. The imaginary part shows slightly larger deviations than the real part, particularly at lower values of $\Omega_m$, though the overall trends remain closely aligned. This indicates that the Chaplygin gas modifies both the oscillation frequency and the damping rate of blackhole perturbations in a quantitatively comparable manner, with a marginally stronger impact on the damping rate. Interestingly, as seen by comparing, the relative deviations from GR in the QNM frequencies for scalar and electromagnetic perturbations are visually and quantitatively indistinguishable. This suggests that, within the MGCG blackhole framework and for the $l = 2$ fundamental mode, the relative departure from the Schwarzschild QNM spectrum is largely independent of the spin of the perturbing field.
Furthermore, we find that Fig.~(\ref{normalisedQNMplot}) the decay rate is only weakly dependent on the multipole number $l$. The imaginary part of $\omega$ shows minimal variation with $l$, especially in the linear regime. In contrast, the real part of the QNM frequency-corresponding to the oscillation frequency-varies significantly with $l$, showing a clear dependence on the angular momentum of the perturbation.

Fig.(\ref{complexplots}) shows the complex-plane trajectories of fundamental QNM frequencies for scalar perturbations with $l=2,3,4$. The curves rise diagonally also in  Fig.~(\ref{complexplotem}), indicating that higher oscillation frequencies $\text{Re}(\omega)$ are consistently accompanied by stronger damping $\text{Im}(\omega)$.
The matter density parameter $\Omega_m$ shifts the spectra systematically. Smaller $\Omega_m$ places the trajectories at lower real frequencies and damping, while larger $\Omega_m$ moves them to the right and upward, producing higher-frequency, more strongly damped modes. Increasing the multipole number $l$ stretches the trajectories horizontally, raising oscillation frequencies while only modestly affecting damping. Despite these scalings, the curves retain nearly identical shapes across $l$, showing a universal dependence on $\alpha$. A slight convex curvature further reflects the nonlinear influence of the Chaplygin gas parameter.

\section*{Conclusion}\label{conclusion}
The Modified Generalized Chaplygin Gas (MGCG) model has been studied in the context of blackhole spacetimes. It is found that this model leads to an asymptotically non-flat blackhole geometry with two distinct horizons: an event horizon and a cosmological horizon. Closed-form expressions for both horizons have been obtained. Compared to the Schwarzschild-de Sitter blackhole, the MGCG model introduces one additional parameter, yet the horizon structure remains qualitatively unchanged. Importantly, the presence of the $\alpha$ parameter relaxes the lower bound on $\Omega_{m}$ imposed by $\Lambda$CDM models.

Scalar and electromagnetic perturbations of the MGCG blackhole have been analyzed using the WKB method, within the parameter regime where the method remains valid. Regions leading to quasi-resonant or long-lived states-where the WKB method breaks down-have been carefully excluded from the analysis. The effective potentials for scalar and electromagnetic perturbations show only marginal differences, typically in the range of $\mathcal{O}(10^{-2})$ to $\mathcal{O}(10^{-6})$.

The computed quasi-normal modes (QNMs) confirm the stability of the MGCG blackhole under both scalar and electromagnetic perturbations. The decay rates of perturbations have been studied as functions of both $\Omega_{m}$ and $\alpha$. The results reveal that higher multipole numbers ($\ell$) increase the oscillation frequency. However, they also result in a longer stabilization time for the perturbing field. The first overtone modes exhibit faster decay rates Fig.~(\ref{QNMn=1tables}).

In addition, the QNM spectra show significant deviations from the Schwarzschild case in both real and imaginary parts, with the imaginary component slightly more affected. These deviations depend on both $\alpha$ and $\Omega_{m}$ but remain nearly identical for scalar and electromagnetic perturbations, indicating that the departure from the Schwarzschild spectrum is largely independent of the perturbing field. Complex-plane trajectories further demonstrate that increasing $\Omega_{m}$ enhances both oscillation and damping, while increasing $\ell$ mainly raises the oscillation frequency.

In future work, it would be interesting to study the MGCG blackhole in the near-extremal regime and explore the onset of quasi-resonances more closely.
\section*{Acknowledgments}
The author would like to thank Anjan Ananda Sen for suggesting the study of this model and for the valuable discussions during the course of this work. SSB thanks the University Grants Commission of India for financial support under the Senior Research Fellowship (SRF) scheme (Ref. No. 191620070524/2020).

\bibliography{ref}

\bibliographystyle{utphys1}
\end{document}